\documentclass[sigconf,authorversion]{acmart}
\AtBeginDocument{%
  }
    
\usepackage[utf8]{inputenc}

\usepackage[]{hyperref} 
\usepackage{pifont}
\usepackage{wasysym}
\usepackage{makecell}
\usepackage{multirow}
\usepackage[figuresleft]{rotating}
\usepackage{longtable,supertabular,booktabs}
\usepackage{threeparttablex,caption}
\usepackage{subcaption}
\usepackage{upgreek}
\usepackage{ltablex}
\usepackage[nolist]{acronym}

\usepackage{cleveref}
\usepackage{amsthm}
\usepackage{bm}
\usepackage{enumitem}
\usepackage{algorithm}
\usepackage[noend]{algpseudocode}

\usepackage{tikz}
\newcommand*\circled[1]{\tikz[baseline=(char.base)]{
            \node[shape=circle,draw,inner sep=0.4pt, font=\footnotesize] (char) {#1};}}

\usepackage{xspace}
\newcommand{\etal}{et al.\xspace}

\newcommand{\ie}{i.e.,\xspace}
\newcommand{\eg}{e.g.,\xspace}
\newcommand{\cf}{cf.\xspace}

\let\oldnl\nl
\newcommand{\nonl}{\renewcommand{\nl}{\let\nl\oldnl}}

\newcommand{\sys}{\textsl{\mbox{BarkBeetle}}\xspace}

\usepackage[textsize=scriptsize]{todonotes}
\let\oldtodo\todo
\renewcommand{\todo}[1]{\oldtodo[inline]{#1}}

\begin{document}

\title[\sys]{\sys: Stealing Decision Tree Models with Fault Injection}

\author{Qifan Wang}
\affiliation{%
  \institution{University of Birmingham}
  \city{Birmingham}
  \country{UK}}
\affiliation{%
  \institution{Durham University}
  \city{Durham}
  \country{UK}}
\email{qifan.wang2@durham.ac.uk}

\author{Jonas Sander}
\affiliation{%
  \institution{University of Luebeck}
  \city{Luebeck}
  \country{Germany}}
\email{j.sander@uni-luebeck.de}

\author{Minmin Jiang}
\affiliation{%
  \institution{Queen’s University Belfast}
  \city{Belfast}
  \country{UK}}
\email{m.jiang@qub.ac.uk}

\author{Thomas Eisenbarth}
\affiliation{%
  \institution{University of Luebeck}
  \city{Luebeck}
  \country{Germany}}
\email{thomas.eisenbarth@uni-luebeck.de}

\author{David Oswald}
\affiliation{%
  \institution{University of Birmingham}
  \city{Birmingham}
  \country{UK}}
\affiliation{%
  \institution{Durham University}
  \city{Durham}
  \country{UK}}
\email{david.f.oswald@durham.ac.uk}

\begin{acronym}[ASCII]
 \setlength{\itemsep}{0.2em}
 \acro{3DES}{Triple DES}
 \acro{AES}{Advanced Encryption Standard}
 \acro{ANF}{Algebraic Normal Form}
 \acro{API}{Application Programming Interface}
 \acro{ARX}{Addition, Rotate, XOR}
 \acro{ASK}{Amplitude-Shift Keying}
 \acro{ASIC}{Application Specific Integrated Circuit}
 \acro{BL}{Bootloader Enable}
 \acro{BOR}{Brown-Out Reset}
 \acro{BPSK}{Binary Phase Shift Keying}
 \acro{CBC}{Cipher Block Chaining}
 \acro{CBS}{Critical Bootloader Section}
 \acro{CGM}{Continuous Glucose Monitoring System}
 \acro{CMOS}{Complementary Metal Oxide Semiconductor}
 \acro{COPACOBANA}{Cost-Optimized Parallel Code Breaker and Analyzer}
 \acro{CPA}{Correlation Power Analysis}
 \acroplural{CPA}[CPAs]{Correlation Power Analyzes}
 \acro{CPU}{Central Processing Unit}
 \acro{CTR}{Counter \acroextra{(mode of operation)}}
 \acro{CRC}{Cyclic Redundancy Check}
 \acro{CRP}{Code Readout Protection}
 \acro{DES}{Data Encryption Standard}
 \acro{DC}{Direct Current}
 \acro{DDS}{Digital Direct Synthesis}
 \acro{DFT}{Discrete Fourier Transform}
 \acro{DMA}{Direct Memory Access}
 \acro{DoS}{Denial-of-Service}
 \acro{DFA}{Differential Fault Analysis}
 \acro{DPA}{Differential Power Analysis}
 \acro{DRAM}{Dynamic Random-Access Memory}
 \acro{DRM}{Digital Rights Management}
 \acro{DSO}{Digital Storage Oscilloscope}
 \acro{DSP}{Digital Signal Processing}
 \acro{DST}{Digital Signature Transponder}
 \acro{DUT}{Device Under Test}
 \acroplural{DUT}[DUTs]{Devices Under Test}
 \acro{ECB}{Electronic Code Book}
 \acro{ECC}{Elliptic Curve Cryptography}
 \acro{ECU}{Electronic Control Unit}
 \acro{EDE}{Encrypt-Decrypt-Encrypt \acroextra{(mode of operation)}}
 \acro{EEPROM}{Electrically Erasable Programmable Read-Only Memory}
 \acro{EM}{Electro-Magnetic}
 \acro{FFT}{Fast Fourier Transform}
 \acro{FCC}{Federal Communications Commission}
 \acro{FIR}{Finite Impulse Response} 
 \acro{FIVR}{Fully Integrated Voltage Regulator}
 \acro{FPGA}{Field Programmable Gate Array}
 \acro{FSK}{Frequency Shift Keying}
 \acro{GMSK}{Gaussian Minimum Shift Keying}
 \acro{GPIO}{General Purpose I/O}
 \acro{GPU}{Graphics Processing Unit}
 \acro{HD}{Hamming Distance}
 \acro{HDL}{Hardware Description Language}
 \acro{HF}{High Frequency}
 \acro{HMAC}{Hash-based Message Authentication Code}
 \acro{HW}{Hamming Weight}
 \acro{IC}{Integrated Circuit}
 \acro{ID}{Identifier}
 \acro{ISM}{Industrial, Scientific, and Medical \acroextra{(frequencies)}}
 \acro{IIR}{Infinite Impulse Response}
 \acro{IP}{Intellectual Property}
 \acro{IoT}{Internet of Things}
 \acro{IV}{Initialization Vector}
 \acro{JTAG}{Joint Test Action Group}
 \acro{LF}{Low Frequency}
 \acro{LFSR}{Linear Feedback Shift Register}
 \acro{LQI}{Link Quality Indicator}
 \acro{LSB}{Least Significant Bit}
 \acro{LSByte}{Least Significant Byte}
 \acro{LUT}{Look-Up Table}
 \acro{MAC}{Message Authentication Code}
 \acro{MF}{Medium Frequency}
 \acro{MMU}{Memory Management Unit}
 \acro{MITM}{Man-In-The-Middle}
 \acro{MSB}{Most Significant Bit}
 \acro{MSByte}{Most Significant Byte}
 \acro{MSK}{Minimum Shift Keying}
 \acro{MSR}{Model Specific Register}
 \acro{muC}[$\mathrm{\upmu C}$]{Microcontroller}
 \acro{NLFSR}{Non-Linear Feedback Shift Register}
 \acro{NLF}{Non-Linear Function}
 \acro{NFC}{Near Field Communication}
 \acro{NRZ}{Non-Return-to-Zero \acroextra{(encoding)}}
 \acro{NVM}{Non-Volatile Memory}
 \acro{OOK}{On-Off-Keying}
 \acro{OP}{Operational Amplifier}
 \acro{OTP}{One-Time Password}
 \acro{PC}{Personal Computer}
 \acro{PCB}{Printed Circuit Board}
 \acro{PhD}{Patiently hoping for a Degree}
 \acro{PKE}{Passive Keyless Entry}
 \acro{PKES}{Passive Keyless Entry and Start}
 \acro{PKI}{Public Key Infrastructure}
 \acro{PMBus}{Power Management Bus}
 \acro{PoC}{Proof-of-Concept}
 \acro{POR}{Power-On Reset}
 \acro{PPC}{Pulse Pause Coding}
 \acro{PRNG}{Pseudo-Random Number Generator}
 \acro{PSK}{Phase Shift Keying}
 \acro{PWM}{Pulse Width Modulation}
 \acro{RAPL}{Running Average Power Limit}
 \acro{RDP}{Read-out Protection}
 \acro{RF}{Radio Frequency}
 \acro{RFID}{Radio Frequency IDentification}
 \acro{RKE}{Remote Keyless Entry}
 \acro{RNG}{Random Number Generator}
 \acro{ROM}{Read Only Memory}
 \acro{ROP}{Return-Oriented Programming}
 \acro{RSA}{Rivest Shamir and Adleman}
 \acro{SCA}{Side-Channel Analysis}
 \acro{SDR}{Software-Defined Radio}
 \acro{SGX}{Software Guard Extensions}
 \acro{SNR}{Signal to Noise Ratio}
 \acro{SHA}{Secure Hash Algorithm}
 \acro{SHA-1}{Secure Hash Algorithm 1}
 \acro{SHA-256}{Secure Hash Algorithm 2 (256-bit version)}
 \acro{SMA}{SubMiniature version A \acroextra{(connector)}}
 \acro{SMBus}{System Management Bus}
 \acro{I2C}{Inter-Integrated Circuit}
 \acro{SPA}{Simple Power Analysis}
 \acro{SPI}{Serial Peripheral Interface}
 \acro{SPOF}{Single Point of Failure}
 \acro{SoC}{System on Chip}
 \acro{SVID}{Serial Voltage Identification}
 \acro{SWD}{Serial Wire Debug}
 \acro{TCB}{Trusted Computing Base} 
 \acro{TEE}{Trusted Execution Environment}
 \acro{TMTO}{Time-Memory Tradeoff}
 \acro{TMDTO}{Time-Memory-Data Tradeoff}
 \acro{TZ}[TrustZone]{TrustZone}
 \acro{RSSI}{Received Signal Strength Indicator}
 \acro{SHF}{Superhigh Frequency}
 \acro{UART}{Universal Asynchronous Receiver Transmitter}
 \acro{UCODE}[$\mathrm{\upmu Code}$]{Microcode}
 \acro{UHF}{Ultra High Frequency}
 \acro{UID}{Unique Identifier} 
 \acro{USRP}{Universal Software Radio Peripheral}
 \acro{USRP2}{Universal Software Radio Peripheral (version 2)}
 \acro{USB}{Universal Serial Bus} 
 \acro{VHF}{Very High Frequency}
 \acro{VLF}{Very Low Frequency}
 \acro{VHDL}{VHSIC (Very High Speed Integrated Circuit) Hardware Description Language}
 \acro{VR}{Voltage Regulator}
 \acro{WLAN}{Wireless Local Area Network}
 \acro{XOR}{Exclusive OR}
 \acro{IR}{Intermediate Representation}
 \acro{OCD}{On-Chip Debug}
 \acro{OS}{Operating System}
 \acro{VM}{Virtual Machine}
 \acro{ML}{machine learning}
 \acro{DVFS}{Dynamic Voltage and Frequency Scaling}
 \acro{MMIO}{Memory-Mapped I/O}
 \acro{TZASC}{TrustZone Address Space Controller}
 \acro{CC}{Confidential Computing}
 \acro{ML}{Machine Learning}
 \acro{DT}{Decision Tree}
 \acro{NN}{Neural Network}
 \acro{DNN}{Deep Neural Network}
 \acro{MLaaS}{Machine Learning as a Service}
 \acro{XAI}{Explainable Artificial Intelligence}
 \acro{GBDT}{Gradient Boosted Decision Tree}
 \acro{FIA}{Fault Injection Attack}
\end{acronym}

\acused{AES}
\acused{CRC}
\acused{DES}
\acused{EEPROM}
\acused{RSA}
\acused{USB}
\acused{SHA-1}
\acused{IC}
\acused{CPU}
\acused{USB}
\acused{DRAM}

\begin{abstract}
Machine learning (ML) models—particularly decision trees (DTs)—are widely adopted across various domains due to their interpretability and efficiency. 
However, as ML models become increasingly integrated into privacy-sensitive applications, concerns about their confidentiality have grown—particularly in light of emerging threats such as model extraction and fault injection attacks.
Assessing the vulnerability of DTs under such attacks is therefore important.
In this work, we present \sys, a novel model extraction attack that leverages fault injection to recover internal structural information of DT models under black-box settings. 
\sys employs a bottom-up recovery strategy that uses targeted fault injection at specific nodes to efficiently infer feature splits and threshold values. 
Our proof-of-concept implementation demonstrates that \sys requires significantly fewer queries and recovers more structural information compared to prior state-of-the-art approaches, when evaluated on DTs trained with public UCI datasets. 
To validate its practical feasibility, we implement \sys on a Raspberry Pi RP2350 microcontroller and perform fault injections using the Faultier voltage glitching tool.
As \sys targets general DT models, we also provide an in-depth discussion on its applicability to a broader range of tree-based applications, including data stream classification, DT model variants, and tree-based cryptography schemes.
\end{abstract}



\acmConference[ASIA CCS '26]{ACM Asia Conference on Computer and Communications Security}{June 1--5, 2026}{Bangalore, India}
\acmBooktitle{ACM Asia Conference on Computer and Communications Security (ASIA CCS '26), June 1--5, 2026, Bangalore, India}
\acmDOI{10.1145/3779208.3785372}
\acmISBN{979-8-4007-2356-8/26/06}

\maketitle

\section{Introduction}

Recent advancements in \ac{ML}, particularly in \ac{DT} models, have driven significant progress across a variety of applications. 
Due to their strong interpretability, \acp{DT} are widely adopted in areas such as online health diagnostics and data stream mining~\cite{shouman2011using,domingos2000mining}. 
To further facilitate \ac{ML} deployment, cloud-based platforms such as Amazon, Google, Microsoft, and BigML have emerged, offering \ac{MLaaS} \ac{API} for model training and inference.
At the same time, ML is increasingly supported on embedded devices. 
Platforms from Arduino, Adafruit, and STMicroelectronics support lightweight ML frameworks such as TensorFlow Lite for Microcontrollers \cite{david2020tensorflow}, Edge Impulse \cite{hymel2022edge}, and Emlearn \cite{emlearn}, enabling inference on resource-constrained devices.
In privacy-sensitive fields such as healthcare and fraud detection, the models themselves may contain proprietary or sensitive information, potentially revealing secret information about the training data or model’s structure. 
Preserving the confidentiality of ML models is therefore essential, as unauthorized extraction or replication can allow adversaries to bypass detection mechanisms or cause critical failures, particularly in high-stakes applications like medical decision-making.

Recently, the increasing commercialization of \ac{ML} services and growing reliance on third-party models have introduced new security and privacy concerns for organizations. Among these, inference attacks—such as membership inference~\cite{shokri2017membership}, model inversion~\cite{fredrikson2015model}, and model extraction~\cite{tramer2016stealing}—pose significant threats. 
Particularly, model extraction (or say stealing) attacks, first introduced by Tram\`er \etal~\cite{tramer2016stealing}, target \ac{MLaaS} platforms with the aim of reconstructing proprietary models. 
Following this seminal work, substantial research has been devoted to model extraction, especially for \ac{NN} models~\cite{wu2022model,hu2020deepsniffer,jagielski2020high,liang2024model,chandrasekaran2020exploring}, where attackers attempt to approximate the target model under various adversarial goals. 
The strongest form of such attacks seeks to replicate the functionality of the target model with near-identical behavior across all possible inputs.
In contrast, only a limited number of studies have focused on extracting tree-based models~\cite{tramer2016stealing,chandrasekaran2020exploring,oksuz2024autolycus}. These works typically rely on black-box access to the model’s prediction interface and employ techniques such as identifying unique outputs, query synthesis via active learning, or the use of \ac{XAI} methods. However, these approaches often face limitations—including dependency on rich output information, high query complexity, and partial recovery of the model. 
For instance, extracting only decision boundaries while overlooking repeated feature usage within each path. This information can, in fact, reveal deeper insights into the feature’s relative importance, its influence on decision-making, and even the underlying distribution of the training data.
Such detailed leakage is essential to facilitate membership inference~\cite{shokri2017membership}, training data reconstruction/model inversion~\cite{fredrikson2015model}, and adversarial example crafting~\cite{carlini2017towards}, and it also allows an adversary to audit the extracted tree for fairness violations, shortcut rules, or improper feature usage~\cite{park2022fairness}—issues that may be invisible from black-box predictions alone.

Another emerging area of interest involves the use of \acp{FIA} to induce faults in targeted regions of a \ac{ML} model, with the aim of generating controlled erroneous outputs and, ultimately, revealing the model’s structure or internal parameters. 
Prior research in this space has primarily focused on \ac{NN} models, which can be broadly categorized into two types of attacks: \textit{misclassification}~\cite{liu2017fault,breier2018practical,hou2021physical,hong2019terminal} and \textit{parameter recovery}~\cite{breier2021sniff,rakin2022deepsteal,hector2023fault}. Misclassification attacks aim to degrade model accuracy by introducing minimal bit flips to the model parameters. 
In contrast, parameter recovery attacks seek to extract internal components of the model, such as weights and biases, particularly in the last layer.
Despite the extensive attention given to \acp{NN}, the application of fault injection techniques to \ac{DT} models remains unexplored—especially in terms of extracting both structural and statistical information. 
In the context of misclassification, \acp{DT} are often more vulnerable to faults due to their simple and deterministic structure, which makes it easier for injected faults to significantly impact prediction accuracy.
Furthermore, most existing works in \acp{FIA} have focused on theoretical feasibility, with limited realization of practical \acp{FIA} on devices.

Recently, \ac{DT} models have demonstrated notable utility in various domains, including data stream classification~\cite{domingos2000mining}, ensemble learning methods such as \ac{GBDT} \cite{friedman2001greedy} and XGBoost \cite{chen2016xgboost}, and privacy-preserving systems such as zero-knowledge proofs and tree-based cryptographic constructions~\cite{mondal2024zkfault,boyle2016function,zhang2020zero}. 
Given their increasing adoption in both performance-critical and security-sensitive applications, it is essential to assess the confidentiality of \acp{DT} under a range of adversarial threats—particularly model extraction and \acp{FIA}.
Note that even cryptographic schemes~\cite{mondal2024zkfault,boyle2016function,zhang2020zero} can be vulnerable to \acp{FIA} (\cf Section~\ref{sec:discuss_tree_crypto}).
Motivated by this gap, we propose \sys, a novel attack that integrates fault injection in the model extraction pipeline to effectively recover the full structure of a tree model. 
\sys performs a bottom-up reconstruction of the tree—recovering nodes from the leaves to the root. 
The attacker injects faults at specific internal nodes to manipulate branching behavior and observe resulting prediction labels. 
By selectively forcing the model to traverse left or right subtrees, the attacker can infer the features and thresholds used at each node.
Our objective is to recover all internal nodes and associated decision tests (\ie features and threshold values) across all paths of a \ac{DT}, while minimizing the number of required queries. 

\subsection{Contributions}
The key contributions of this work are summarized as follows:
\begin{itemize}[noitemsep,topsep=0pt]
    \item To the best of our knowledge, \sys is the first attack to apply fault injection for extracting decision tree models. As it targets general tree structures, \sys also provides a foundation for extending it to other widely used tree-based models in various settings (\cf Section~\ref{sec:limitation_discussion}).
    \item We design a bottom-up algorithm that leverages fault injection at target nodes to recover features and thresholds more efficiently. This approach reduces the number of required queries and is particularly effective for tree-based models, which share a hierarchical decision-making structure.
    \item We implement a proof-of-concept attack for \sys and show that \sys significantly reduces the number of queries compared to prior relevant work~\cite{tramer2016stealing} on \acp{DT} trained with UCI datasets. In addition, \sys reveals more information, including duplicate features along each path, providing deeper insight into their influence and significance within the decision process.
    \item To demonstrate the practical feasibility of our attack, we evaluate \sys on a Raspberry Pi RP2350 board and implement the voltage glitch using the Faultier tool \cite{faultier}. In our case study with a \ac{DT} (depth 5, 11 leaves, trained on the Diabetes Diagnosis dataset), the attack required 703 additional queries due to repeated glitch attempts. This overhead could be reduced by employing more precise fault injection techniques.
    \item To support reproducibility and future research, we will release the source code for \sys upon acceptance, enabling the development and evaluation of countermeasures. For the reviewers, an anonymised version is available at \url{https://github.com/DylanWangWQF/BarkBeetle}. 
\end{itemize}

\section{Background and Related Work}
In this section, we first present the basics of \ac{DT} models, followed by a summary of existing research on model extraction and \acp{FIA} in \ac{ML}. Additionally, we introduce some commonly used \ac{FIA} techniques.

\subsection{Decision Tree}
\label{sec:dt basics}
A \ac{DT} consists of internal nodes and leaves, where each internal node is associated with a test on a feature, each branch represents the outcome of the test, and each leaf represents a label which is the decision taken after testing all the features on the corresponding path.
The label value of the accessed leaf is the inference result.

The splits in a \ac{DT} are determined by the features, which can be either continuous or categorical. 
Continuous features represent numerical values, such as \textit{age} or \textit{temperature}, and are split using threshold values (\eg $age<20$). 
In contrast, categorical features represent distinct classes (\eg color) and are split based on categories (\eg red and blue resulting in two branches).

Some \ac{ML} prediction \acp{API}, such as Amazon\footnote{\url{https://aws.amazon.com/ machine-learning}} and BigML\footnote{\url{https://bigml.com/api/}}, return information-rich outputs—specifically, high-precision \textit{confidence scores} in addition to class labels. 
Specifically, each leaf is assigned both a class label and a confidence score. 
These prediction labels and confidence values can serve as unique identifiers for nodes, allowing an attacker to distinguish between leaves that share the same label, as demonstrated in~\cite{tramer2016stealing}.

\subsubsection{Classification Tree and Regression Tree}

A classification tree is typically used for categorical target labels, assigning inputs to predefined classes (\eg ``Yes" or ``No" for predicting whether to play tennis under different weather conditions). 
A regression tree, on the other hand, is designed for continuous labels, predicting numerical values such as house prices or temperatures. Instead of classifying, it partitions the data to minimize variance or mean squared error in the target variable within each group \cite{quinlan2014c4}.

A key factor influencing \ac{DT} extraction is the difference in output values between classification and regression trees. 
While classification trees predict discrete categories, regression trees output continuous values. 
As a result, different paths in a classification tree may lead to the same label, whereas all paths in a regression tree typically yield unique outputs.
However, when multiple paths in a classification tree share the same label, it can reduce the attacker’s ability to distinguish between different paths (\cf Section~\ref{sec:discuss_same_label}).

\subsection{Model Extraction Attack on \ac{ML} Models}
\label{sec:related_mea}
Model extraction attacks pose a serious threat to \ac{MLaaS} platforms by stealing the functionality of private \ac{ML} models through querying black-box \acp{API}.
Since the pioneering study by~\cite{tramer2016stealing}, research on these attacks has primarily focused on \ac{NN} models~\cite{wu2022model,hu2020deepsniffer,jagielski2020high,liang2024model,chandrasekaran2020exploring}.
These attacks consists of three key steps:
i) generating queries, either sampled from the training dataset or synthesized;
ii) sending each query to the \ac{MLaaS} \ac{API} and receiving its prediction result;
iii) iteratively updating the stolen model using the query-response pairs~\cite{liang2024model}.
Adversarial goals in model extraction attacks can be classified into three categories:
i) \textit{Functionally Equivalent Extraction}: The attacker creates a model that behaves identically to the target for all inputs (a strong, input-agnostic goal).
ii) \textit{Fidelity Extraction}: The attacker extracts a model that closely matches the target’s outputs on the data distribution of interest, even if it is not identical everywhere.
iii) \textit{Task Accuracy Extraction}: The attacker extracts a model that matches or exceeds the target model’s accuracy. This goal is easier to achieve since the extracted model does not need to replicate the target model’s mistakes~\cite{jagielski2020high}.
Beyond that, query complexity is the most common metric in quantifying the efficiency of attacks. 
Reducing query count allows the attack to scale to large models and can be helpful to overcome countermeasures based on rate limiting or anomaly detection.

Only a few works~\cite{tramer2016stealing,chandrasekaran2020exploring,oksuz2024autolycus} have studied extraction attacks on \ac{DT} models. 
In~\cite{tramer2016stealing}, the authors initiate their attack by submitting a random input $\bm{X}$, obtaining a leaf identifier, and subsequently identifying all constraints on $\bm{X}$ required to remain in that leaf. 
They iteratively generate new queries targeting previously unvisited leaves until the entire tree structure is identified.
The number of queries can be significantly improved by employing incomplete queries—a functionality available through platforms like BigML—where certain input features are intentionally left unspecified. 
Specifically, the model traversal continues until an internal node with a split over a missing feature is reached, at which point the node’s threshold value is directly revealed.
However, it can be easily mitigated by disabling it.
The following work \cite{chandrasekaran2020exploring} relaxes assumptions regarding the availability of rich information, such as leaf identifiers and incomplete queries. 
Instead, they employ the IWAL algorithm introduced by \cite{beygelzimer2010agnostic}, enabling \ac{DT} extraction attacks without relying on auxiliary information.
Finally, their method requires more queries but achieves better accuracy on \acp{DT} containing paths with the same labels.
More recently, \cite{oksuz2024autolycus} exploits \ac{XAI} to target interpretable models, aiming to infer decision boundaries and create surrogate models. 
Their experimental results demonstrate significantly greater efficiency compared to the prior two studies. 
However, it strongly depends on the availability of rich information provided by the \ac{MLaaS} platform—information that may not always be accessible and could be restricted.
Moreover, some of this rich information is part of the attack objectives in \ac{DT} extraction.

Overall, these existing approaches suffer from one or more of the following limitations: 
\ding{182} reliance on rich auxiliary information;
\ding{183} requiring a large number of queries;
\ding{184} recovering only the decision boundaries of features in each tree path while overlooking the feature importance within each path (\ie repeated feature usage within each path).
Specifically, if a feature appears multiple times along a path, it reflects not only its overall importance but also its conditional relevance—its continued predictive relevance following earlier splits.
This pattern can also offer insights into the underlying distribution of the training data \cite{molnar2020interpretable}.
If a feature appears multiple times along a path, it reflects not only its overall importance but also its conditional relevance—that is, its continued predictive relevance following earlier splits. 
This pattern may also suggest interactions or correlations between features, where one feature becomes more informative within specific partitions defined by another~\cite{quinlan2014c4}.
Consistent with prior work showing that internal structural information  strengthen attacks~\cite{liu2022membership,yeom2018privacy,wu2022model}, we can reasonably assume that an attacker with access to such detailed structural information can infer not only the \ac{DT}’s decision logic but also approximations of the underlying training data distribution.

\subsection{Fault Injection Attack on \ac{ML} Models}

\acp{FIA} aim to manipulate a system’s behavior or extract sensitive data by intentionally introducing faults. In this section, we provide an overview of commonly used \ac{FIA} techniques and summarize existing research on \acp{FIA} in \ac{ML}.

\subsubsection{Fault Injection Methods}
Fault injection can be performed using various techniques and equipment, depending on factors such as the number of fault injection events, the fault location, timing, the number of affected bits, the duration of the fault, the types of gates targeted, and the nature of the fault~\cite{Bar2006sorcerer,toprakhisar2024sok}.

\noindent\textbf{Clock/voltage glitch.} 
Voltage glitching is achieved by momentarily dropping the supply voltage during specific operations, while clock glitching disrupts clock timing to violate the hardware’s setup and hold requirements, typically by inserting glitch pulses between normal clock cycles.

\noindent\textbf{\ac{EM} fault injection.} 
An \ac{EM} fault injection involves generating a localized, short-duration, high-intensity electromagnetic pulse, inducing unintended currents in the chip’s internal circuitry.

\noindent\textbf{Optical fault injection.} 
This method uses equipment with varying precision, ranging from camera flashes to lasers. It offers high reproducibility and precision, with lasers capable of inducing exact bit flips.

\noindent\textbf{Rowhammer.} 
The Rowhammer attack is a software-based fault injection technique that can be triggered remotely. 
By repeatedly accessing one memory row, interference with neighboring rows can occur due to the dense memory structure, causing faster charge leakage. 
If the refresh rate is insufficient, bit flips may occur.

Since voltage glitches can be fine-tuned to target specific timing in the execution cycle, allowing precise disruption of specific operations, and are low-cost compared to advanced fault injection methods like laser or optical techniques, we focus on using the voltage glitch method to induce faults in our practical case study (\cf detailed experiment setup in Section~\ref{sec:glitch_exp}).

\subsubsection{Fault Injection Attacks on \ac{ML} Models}
Over the past few decades, extensive research has been conducted on \acp{FIA} targeting \ac{ML} models, particularly \acp{NN}. We summarize this line of research in two main categories based on their attack goals: namely \textit{misclassification} and \textit{parameter recovery}.
In misclassification-based studies~\cite{liu2017fault,breier2018practical,hou2021physical,hong2019terminal}, the goal is to reduce the model’s accuracy drastically by introducing as few faulty bits as possible into the target \ac{NN}’s parameters. 
For example, \cite{hong2019terminal} examined the effects of bitwise corruptions on 19 \ac{DNN} models across six architectures and three image classification tasks. 
They found that factors such as the bit-flip position, flip direction, parameter sign, and layer width have a significant impact on the extent of the accuracy degradation.

In parameter recovery-based studies~\cite{breier2021sniff,rakin2022deepsteal,hector2023fault}, the objective is to partly recover \ac{NN} parameters, such as the weights and biases in the last layer. 
For instance, \cite{hong2019terminal} employed the Rowhammer attack to recover significant bits (from LSB to MSB) and train a substitute model. 
\cite{hector2023fault} extended this approach to a 32-bit microcontroller, recovering over $90\%$ of the most significant bits using around 1500 crafted inputs, allowing them to train a near-equivalent model.

Despite extensive research on fault injection in \acp{NN}, no existing studies have explored the use of \acp{FIA} to extract both the structural and statistical information of \ac{DT} models. 
In terms of misclassification in \ac{DT} models, this is likely because the structural characteristics of \acp{DT} make it relatively easy to induce faults which can significantly degrade the model accuracy compared to \acp{NN}.
In this work, we explore \acp{FIA} in \ac{DT} models with the goal of extracting (or say recovering) both their structural and statistical information.

\section{Data and Model Representation}
\label{sec:dm_rep}

\begin{table}[tb]
    \centering
    \footnotesize
    \caption{Notations used in this paper.}
    \label{tab:notation}
    \begin{tabular}{c l}
        \toprule
        \textbf{Notation} & \textbf{Description} \\ 
        \midrule
        $d$ & Number of features \\
        $\bm{X}$ & Data sample where $|\bm{X}|=d$ and $\bm{X}=(x_0,x_1,{\cdots},x_{d-1})$ \\
        $\bm{S}$ & Feature set, $\bm{S}=(s_0,s_1,{\cdots},s_{d-1})$ \\
        $\bm{V}$ & \makecell[l]{Tree node, $\bm{V}=(s,t,br)$ where $s$, $t$, and $br$ denote the feature,\\ threshold values,and edge direction in this node} \\ 
        $\bm{L}$ & Labels in tree leafs, $\bm{L}=(c_0,c_1,{\cdots},c_{\alpha-1})$ \\
        $\bm{P}$ & \makecell[l]{Tree path, $\bm{P}=\{(\bm{V_0},\bm{V_1},{\cdots},\bm{V_{\beta-1}}),c\}$ contains  $\beta$ nodes on\\ the path and label $c$} \\
        $\Upsilon$ & Tree, comprised of $\alpha$ tree paths $(\bm{P_0},\bm{P_1},{\cdots},\bm{P_{\alpha-1}})$  \\
        $t_i^j$ & \makecell[l]{Threshold in the $j$-th (from the leaf to the root) duplicated node\\ for feature $s_i$}\\
        \midrule
        \multicolumn{2}{c}{\textbf{Algorithms}} \\
        \cmidrule{1-2}
        $\mathtt{F\_Inf}()$ & \makecell[l]{Algorithm~\ref{algo_fi}: Fault-based Inference (Normal inference is\\ denoted as $\mathtt{Inf}()$)} \\
        $\mathtt{FaBS}()$ & Algorithm~\ref{algo_bs}: Fault-assisted Binary Search \\
        $\mathtt{FFD}()$ & Algorithm~\ref{algo_ffd}: First Feature Discovery \\
        $\mathtt{DFD}()$ & Algorithm~\ref{algo_dfd}: Duplicate Features Discovery \\
        $\mathtt{RTI}()$ & Algorithm~\ref{algo_rti}: Recover Tree Iterator \\
        $\mathtt{TreeExt}()$ & Algorithm~\ref{algo_general}: DT Extraction \\
        \bottomrule
    \end{tabular}
\end{table}

In this section, we present the formal representations for query data array and \ac{DT} model.
In the rest of this paper, we use notations listed in \autoref{tab:notation}. 

\subsection{Data Representation}
In \sys, we primarily focus on continuous features, as they are more prevalent in common \ac{DT} models.
Moreover, categorical features can be more easily identified by systematically adjusting their values and observing the corresponding label changes.

$\bm{S}=(s_0,s_1,{\cdots},s_{d-1})$ represents a set of $d$ continues features with the range $s_i\in[a,b]$.
The minimal and maximal values of feature $s_i$ are denoted as $s_i.a$ and, $s_i.b$ respectively.
A non-labeled query data is denoted as $\bm{X}=(x_0,x_1,{\cdots},x_{d-1})$.
As in prior work \cite{tramer2016stealing}, we assume that the range $[a,b]$ can be inferred from the background information.
For instance, if a \ac{DT} has been deployed and actively used over time, an attacker can estimate $[a,b]$ for each feature by analyzing historical query patterns and responses.

\subsection{Model Representation}
\label{sec:model rep}

Formally, we assume that each internal node is assigned a feature $s_i \in \bm{S}$ and a binary splitting function, \ie $x_i < t_i$, where $x_i$ is the value of the $i$-th feature $s_i$ in the given data sample and $t_i$ is the threshold value for the $i$-th feature in this node determined during training.
A tree $\Upsilon$ consists of a set of paths $(\bm{P_0},\bm{P_1},{\cdots},\bm{P_{\alpha-1}})$, where ${\alpha}$ is the number of paths in the tree.
Each tree path $\bm{P}=(\bm{V_0},\bm{V_1},{\cdots},\bm{V_{\beta-1}}, c)$ contains $\beta$ internal nodes, namely $\bm{V}=(s,t)$, and the corresponding label $c$. 
Without loss of generality, we introduce an auxiliary attribute $br$, to indicate the direction of the edge: $br=0$ indicates that the next node is its left child, while $br=1$ signifies the right child.
This attribute $br$ is essential for crafting inputs that lead to different labels in our attack (\cf Algorithm~\ref{algo_rti}).
Finally, we represent a node as $\bm{V}=(s,t,br)$.
Note that a feature may appear multiple times within the same path or across different paths, as a continuous feature can be split multiple times in the tree.
The labels in the leafs of the \ac{DT} are represented as $\bm{L}=(c_0,c_1,{\cdots},c_{\alpha-1})$.

Given an input $\bm{X}=(x_0,x_1,{\cdots},x_{d-1})$, the \ac{DT}-inference proceeds as follows: Starting at the root, when an internal node $\bm{V}$ is reached, the split function $x_i < \bm{V.t}$ is evaluated.
If $x_i < \bm{V.t}$, the process moves to the left child; otherwise, it moves to the right child.
Once a leaf node is reached, the inference process terminates, and the label stored in the leaf is returned as the result.
We assume that each leaf node has a unique identifier (\eg label).
This is also the reason why \cite{tramer2016stealing} and \sys perform better on regression trees, where each output is a distinct real value, making it easier to use as a unique identifier.
For a discussion on how this restriction can be relaxed, please refer to Section~\ref{sec:discuss_same_label}.

\begin{figure}[tb]
\centering
\includegraphics[width=\linewidth]{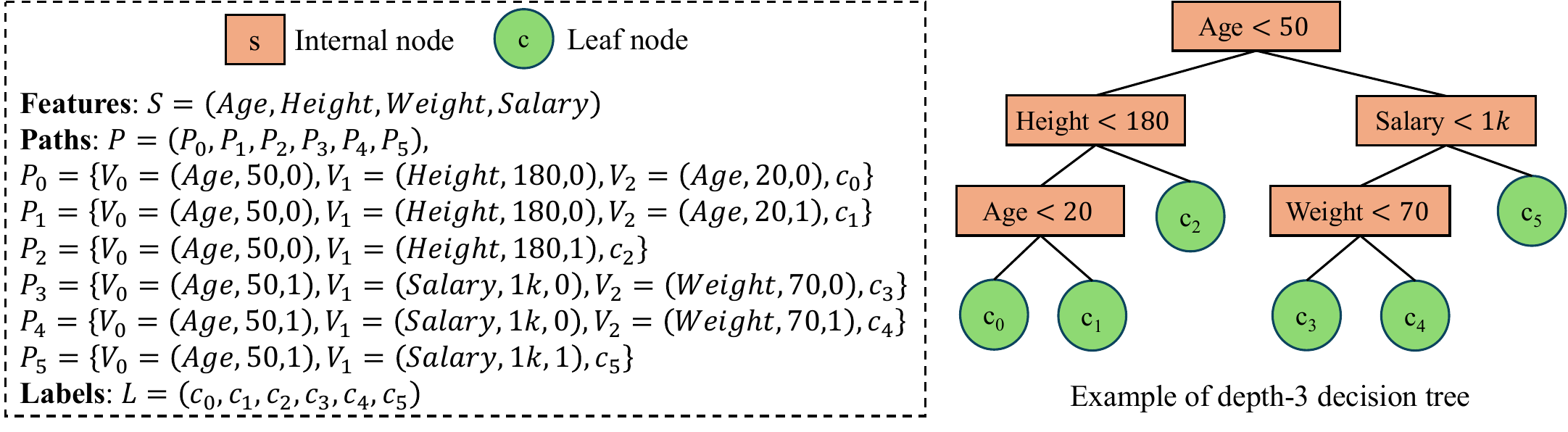}
\caption{Representation of a toy \ac{DT} model.} 
\label{fig:dt_rep}
\end{figure}

We provide a toy example of a depth-3 \ac{DT} in \autoref{fig:dt_rep}. 
Based on our data and model representations, the tree can be represented as feature set $\bm{S}$, path set $\bm{P}$, and label set $\bm{L}$.

\section{Methodology}

\begin{figure*}[htbp]
\centering\includegraphics[width=0.75\linewidth]{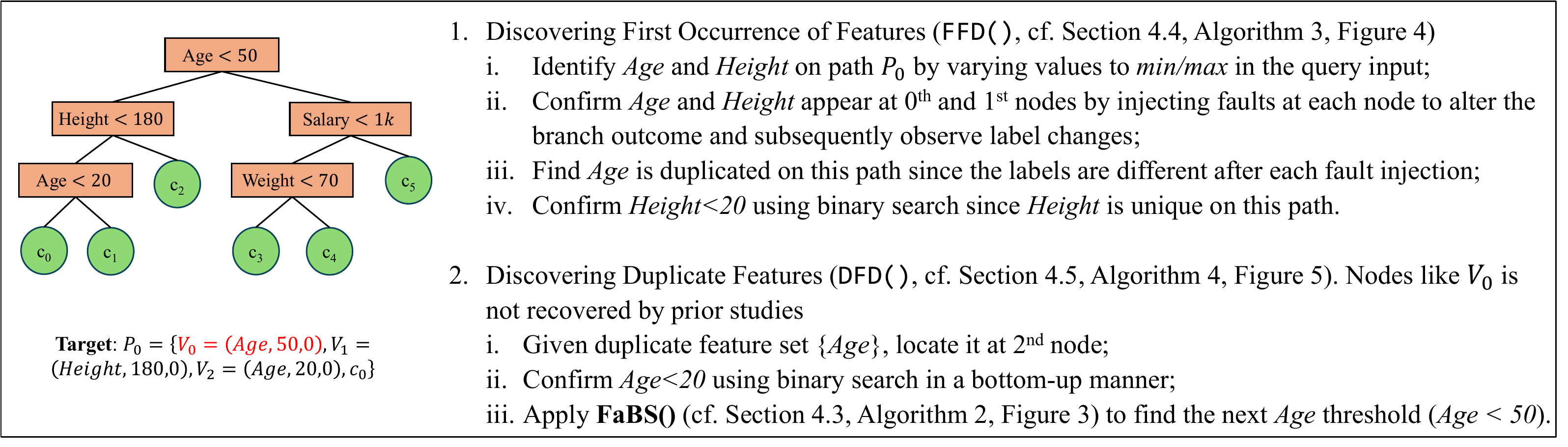}
\caption{Walkthrough example of \sys, illustrating feature recovery steps based on \autoref{fig:dt_rep}.} 
\label{fig:walkthrough}
\end{figure*}

In this section, we begin by presenting the threat model, outlining the attack objectives, capabilities, and knowledge assumed for the adversary in \sys.
We then introduce the building blocks of our attack, explaining the types of faults used in \ac{DT} inference. 
Next, we describe how \sys recovers feature and threshold values by first identifying the initial occurrence of each feature on a path (Algorithm~\ref{algo_ffd}), followed by detecting and recovering duplicate features along the same path (Algorithm~\ref{algo_dfd}).
To aid understanding, \autoref{fig:walkthrough} provides a step-by-step example aligned with \autoref{fig:dt_rep}, illustrating the entire recovery process and how each component fits together.
Finally, we discuss how certain assumptions can be relaxed and outline potential directions for improvement.

\subsection{Threat Model}
\label{sec:threat model}

In this work, we consider the \ac{DT} model $\Upsilon$, which is trained on a dataset and deployed for inference.
We assume that $\Upsilon$ is pre-trained, and any information from the training phase is beyond the scope of this study.
Our work focuses on adversaries targeting \ac{DT} models on embedded devices such as microcontrollers (\eg Raspberry Pi RP2350 microcontroller used in our experiment, \cf Section~\ref{sec:hw_setup}).
This choice is driven by \sys’s use of voltage glitching to induce fault injection, which is well-suited for embedded platforms.
However, importantly, our attack is also applicable to cloud environments if the attacker can perform fault injection attacks within the cloud infrastructure (\eg Rowhammer~\cite{mutlu2019rowhammer}, undervolting~\cite{murdock2019plundervolt,Kenjar20,Qiu19_sgx}, binary code modification~\cite{cotroneo2016faultprog}, compiler-level fault injection~\cite{wei2014quantifying}, or debugger-based techniques~\cite{fang2014gpu}).

While \sys requires physical access for glitch-based injection, such access is realistic in many modern deployments. Lightweight IoT and embedded devices often make physical access a practical threat. Moreover, widely used embedded learning frameworks, \eg Edge Impulse~\cite{hymel2022edge} and Emlearn~\cite{emlearn}, already support tree-based models.

\noindent\textbf{Attack goals.}
Our focus is on an attacker $\mathcal{A}$ who aims to recover the tree structure and node information of $\Upsilon$, \ie \textit{path set} $\bm{P}$ and \textit{label set} $\bm{L}$ (see examples in \autoref{fig:dt_rep}), in a \ac{DT}-based inference system. 
We consider $\mathcal{A}$’s capabilities and knowledge regarding the victim model as follows:

\noindent\textbf{Capabilities.}
First, $\mathcal{A}$ has unlimited black-box access to query the model.
Second, $\mathcal{A}$ can induce controlled, transient faults at specific timing points by briefly varying the supply voltage beyond its nominal range.
Third, $\mathcal{A}$ can exploit measurable variations to infer the number of nodes along each path and estimate the execution windows of individual nodes at each internal node during each inference. 
The latter is required for performing fault injection at target area.
Prior related tree-based cryptographic scheme, such as \cite{mondal2024zkfault}, assumes that the faulted location of the tree node is arbitrary but \textit{known to the attacker}.
In this work, we assume that $\mathcal{A}$ can leverage side-channel signals (\eg power or timing) to estimate a node's execution window and synchronize glitches accordingly~\cite{jap2020practical,batina2019csi,coqueret2024hard,weerasena2024revealing}.
When considering other \ac{FIA} techniques such as Rowhammer, memory-profiling techniques can achieve similar synchronization.
This part is well-studied and orthogonal to to \sys.

However, if consider a weaker grey-box setting as in \cite{hector2023fault}--which is stronger than white-box setting~\cite{hong2019terminal,rakin2022deepsteal}) but weaker than black-box setting--$\mathcal{A}$ may know the number of nodes in each path and the timing duration per node. 
This is common when the model structure is known, easy to infer, or partially revealed by earlier attacks.
In such cases, $\mathcal{A}$ does not require the above third capability of leveraging timing variations.

\noindent\textbf{Knowledge.}
In \sys, $\mathcal{A}$ has no prior knowledge of the victim \ac{DT}'s structure and node information.

\subsection{Fault-based \ac{DT} Inference}

We demonstrate how previously assumed side-channel information is utilized to determine the necessary timing for inducing faults and then introduce the fault types required in \sys.

The attacker $\mathcal{A}$ requires timing information to inject faults at specific tree nodes. We assume that $\mathcal{A}$ can exploit side-channel leakage~\cite{jap2020practical,batina2019csi} to estimate the start and end times of each node’s execution—denoted as $\mathcal{T}[i][0]$ and $\mathcal{T}[i][1]$ for the $i$-th node—via a procedure $\bm{P}, \mathcal{T} \gets \mathtt{CTN}(\zeta_{sc}, \bm{P})$, where $\zeta_{sc}$ represents the observed side-channel signals.

\begin{algorithm}[htbp]
\renewcommand{\algorithmicrequire}{\textbf{Input:}}
\renewcommand{\algorithmicensure}{\textbf{Output:}}
\footnotesize
\caption{Fault-based Inference: $\mathtt{F\_Inf}()$}
\label{algo_fi}
\begin{algorithmic}[1]
\Require Input $\bm{X}$, timing array $\mathcal{T}$ of $\bm{P}$, target node index $idx$; $flag=0/1$ indicates left/right subtree
\Ensure Label $c$ of faulted inference

\If{$flag = 0$}
    \State Flip comparison result of $idx$-th node to make it go left during $[\mathcal{T}_{idx}[0], \mathcal{T}_{idx}[1]]$
\Else
    \State Flip comparison result of $idx$-th node to make it go right during $[\mathcal{T}_{idx}[0], \mathcal{T}_{idx}[1]]$
\EndIf

\State $c \gets \mathtt{Inf}(\bm{X})$
\end{algorithmic}
\end{algorithm}

\begin{figure*}[htbp]
\centering
\includegraphics[width=0.75\linewidth]{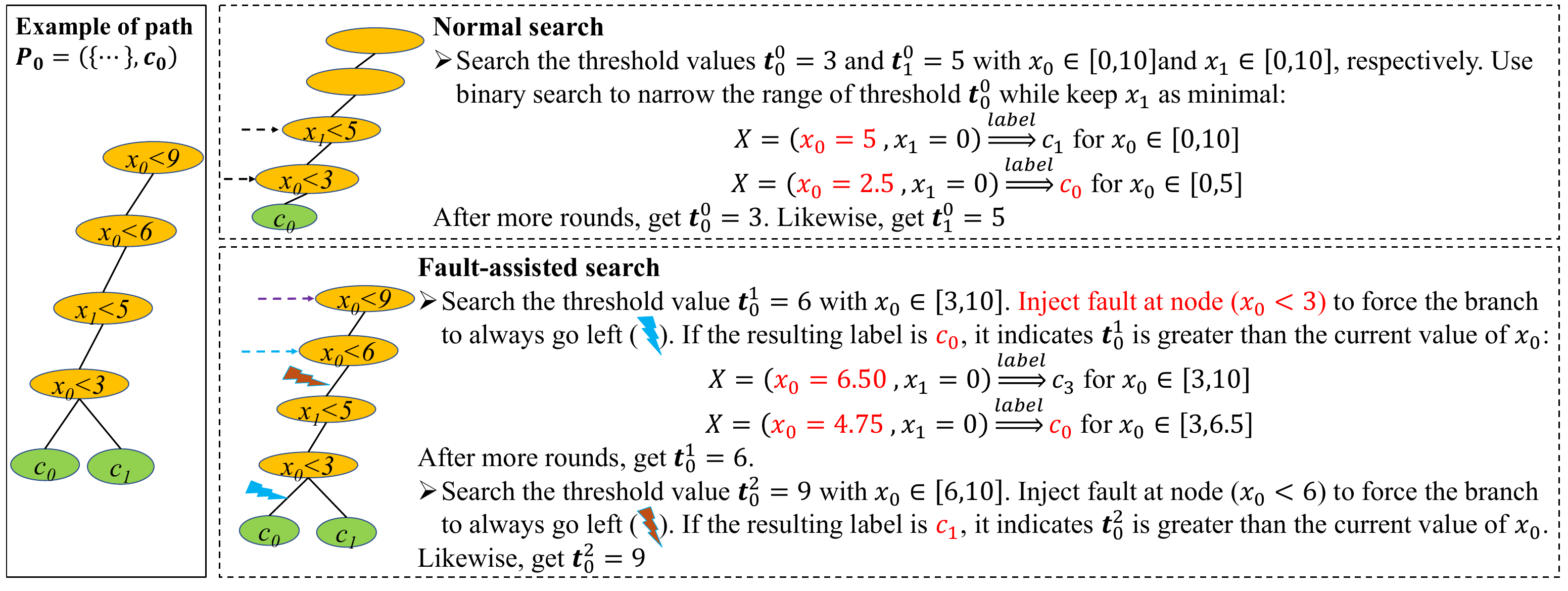}
\caption{Illustration of fault-assisted binary search.} 
\label{fig:bs_example}
\end{figure*}

The core idea of \sys is to manipulate the comparison result at a specific node, enabling the attacker to gradually infer the feature and threshold values on that path.
Algorithm~\ref{algo_fi} describes the types of faults required to achieve this.
In general, \sys injects faults to control branch execution. 
Specifically, in the left subtree (\ie the left partition of the root node), a fault is induced at a node to force the branch to go left, regardless of the original decision. 
For example, given the comparison $x_0<20$ with a feature value of $x_0=25$, the injected fault alters the expected execution flow, forcing the branch to take the left path instead of the originally intended right path.
Conversely, in the right subtree, faults are applied to ensure the branch always goes right.
Finally, $\mathcal{A}$ obtains the label from fault-injected inference.

\subsection{Fault-assisted Binary Search}

\begin{algorithm}[!t]
\renewcommand{\algorithmicrequire}{\textbf{Input:}}
\renewcommand{\algorithmicensure}{\textbf{Output:}}
\footnotesize
\caption{Fault-assisted Binary Search: $\mathtt{FaBS}()$}
\label{algo_bs}
\begin{algorithmic}[1]
\Require Input $\bm{X}$, baseline label $c_b$, timing array $\mathcal{T}$ of $\bm{P}$, feature index $idx$, node index $nidx$, low boundary $low\_bnd$, high boundary $high\_bnd$, flag $ifFIA$, left/right subtree flag $flag$, granularity $\epsilon$
\Ensure Threshold value $t$

\State $c_{ls} \gets c_b$; $low \gets low\_bnd$; $high \gets high\_bnd$

\While{$high - low \geq \epsilon$}
    \State $\bm{X}[idx] \gets (low + high) / 2$
    \If{$ifFIA$}
        \State $c_{ls} \gets \mathtt{F\_Inf}(\bm{X}, \mathcal{T}, nidx, flag)$
    \Else
        \State $c_{ls} \gets \mathtt{Inf}(\bm{X})$
    \EndIf

    \If{($flag = 0$ and $c_{ls} = c_b$) or ($flag = 1$ and $c_{ls} = c_b$)}
        \State $low \gets \bm{X}[idx]$
    \Else
        \State $high \gets \bm{X}[idx]$
    \EndIf
\EndWhile

\State $t \gets \bm{X}[idx]$
\end{algorithmic}
\end{algorithm}

In Algorithm~\ref{algo_bs}, our goal is to determine the threshold values for each feature.
\cite{tramer2016stealing} employs \textit{binary search} to identify the feature value range in which an input, while keeping all other feature values unchanged, leads to the same leaf node.
However, this method failed to recover the repeated feature usage within each path.
For example, consider a path with three nodes: $x<a$, $x<b$, $x>c$ (where $a>b$).
The approach in \cite{tramer2016stealing} would identify the range $c<x<b$ for feature $x$ on this path while neglecting $x<a$ and some nodes.
Our objective is to recover all threshold values, including those for duplicate features appearing at different nodes along the same path.

We propose a fault-assisted binary search, which consists of two types of searches: normal search and fault-assisted search. We use $t_i^j$ to reference the threshold value in $j$-th (from the leaves to the root node) duplicated node regarding feature $s_i$.
\textbf{Normal search}: given lower and upper bounds, we apply binary search to determine the threshold $t$, ensuring that for any input satisfying $x_i < t$ (or $x_i > t$), the traversal leads to the existing leaf.
As illustrated in \autoref{fig:bs_example}, we use binary search to progressively narrow the range of $x_0$ and determine the threshold $t^0_0=3$.
\textbf{Fault-assisted search}: when duplicate features appear along the same path, injecting a fault at one node is insufficient. The key challenge is eliminating the influence of earlier nodes that also involve the same feature. 
To address this, we inject faults at the preceding node to manipulate branch execution and compare the resulting label with the baseline label (as detailed in Algorithm~\ref{algo_bs}). 
This allows us to extend the search range and accurately recover the next threshold value for the feature.
For example, in \autoref{fig:bs_example}, we confirm the thresholds $t_0^1=6$ and $t_0^2=9$ for $x_0$ by injecting faults at earlier nodes (\ie forcing the comparisons at $x_0<3$ and $x_0<6$ to always branch left, respectively).

\subsection{Discovering First Occurrence of Features}
\label{sec:ffd}

\begin{figure*}[htbp]
\centering
\includegraphics[width=0.75\linewidth]{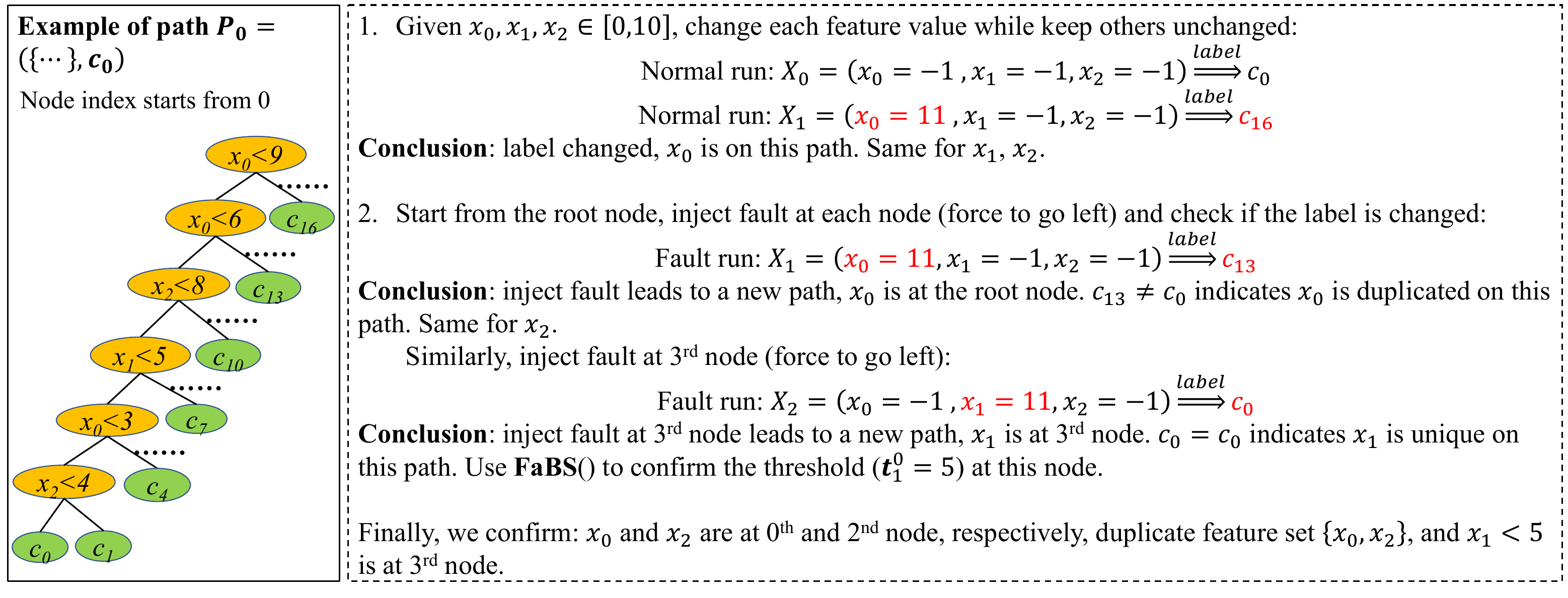}
\caption{Example of $\mathtt{FFD}()$ for recovering the features first appear on this path.} 
\label{fig:ffd_example}
\end{figure*}

Each path may contain both unique features (appearing only once) and duplicate features (appearing multiple times). 
Recovering duplicate features is inherently more complex than identifying unique ones. 
In Algorithm~\ref{algo_ffd} (\cref{appendix:algos}), we first recover the features that appear for the first time along a path, which in turn reduces the complexity of the subsequent step of recovering duplicates by narrowing the search space.
Overall, Algorithm~\ref{algo_ffd} can be summarized as two main steps: 
\ding{182} Modify each feature $x_i$ to $s_i.b+1$ (or $s_i.a-1$ in the right subtree) while keeping all other feature values as $s_i.a-1$, then check whether the label changes. 
For instance, in \autoref{fig:ffd_example}, changing $x_0$ to $s_0.b+1=11$ results in a label change to $c_{16}$, indicating that $x_0$ is present on this path.
\ding{183} We then determine whether it appears once or multiple times. If it is a duplicate, we update the duplicate feature set $S_{DF}$, which will be processed in Algorithm~\ref{algo_dfd}.
For example, when changing $x_1$ to $s_1.b+1$, in \autoref{fig:ffd_example}, injecting a fault at the third node causes the label to change to $c_0$, confirming that $x_1$ is present at this node as a unique feature, allowing us to determine its threshold value ($t^0_1=5$).

\subsection{Discovering Duplicate Features}
\label{sec:dfd}

\begin{figure*}[htbp]
\centering
\includegraphics[width=0.75\linewidth]{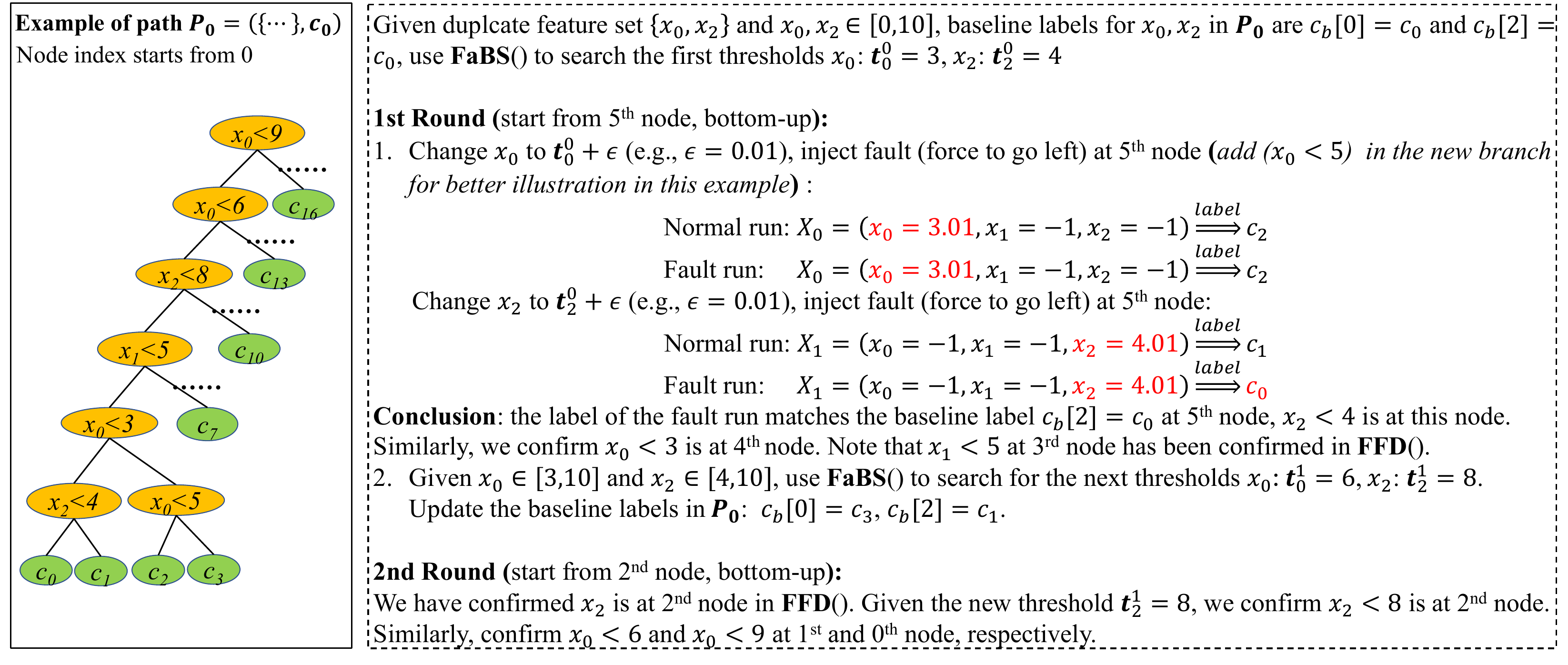}
\caption{Example of $\mathtt{DFD}()$ for recovering duplicate features on this path.}
\label{fig:dfd_example}
\end{figure*}

The primary challenge in recovering nodes with duplicate features along a path is also mitigating the influence of other nodes containing the same feature, as modifying the feature value can unintentionally lead to new paths, complicating the recovery process. 
Algorithm~\ref{algo_ffd} has already narrowed down the set of duplicate features. 
Building on this, we propose a bottom-up algorithm that reconstructs duplicate features from the last internal node to the root node.
The key idea behind this is to recover features starting from the smallest threshold (in the left subtree), thereby preventing interference from nodes that appear earlier in the path.

The process consists of three main steps: 
\ding{182} \textbf{Locating target node}: Given an already confirmed threshold, identify the node that contains this feature and threshold.
To do this, modify $\bm{X}[i]$ to $t_i \pm \epsilon$, ensuring that the modified input leads to a label different from the original label (say baseline label).
Then, inject faults at each node and check whether the label changes to the baseline label. 
If so, the feature $s_i$ with threshold $t$ is confirmed at this node.
For example, in the first round of \autoref{fig:dfd_example}, after modifying $x_2$ to $t^0_2+\epsilon=4.01$, the label of the fault run matches the baseline label $c_0$, confirming that $x_2<4$ belongs to the last node.
$\epsilon$ represents the granularity used in the binary search.
For features with large integer values, $t_i \pm \epsilon$ may have no effect.
In such cases, $\epsilon$ should be experimentally tuned and replaced with a suitable value.
\ding{183} \textbf{Updating feature value range and $S_{DF}$}: Narrow the feature value range and check if $s_i$ appears in any remaining nodes. If not, remove it from $S_{DF}$ to avoid redundant searches. 
\ding{184} \textbf{Searching for next threshold}: The next threshold for $s_i$ is determined using $\mathtt{FaBS}()$, and the baseline label for each duplicate feature is updated accordingly.
For instance, in \autoref{fig:dfd_example}, after confirming $x_2<4$ and $x_0<3$, the baseline labels for $x_2$ and $x_0$ are updated to $c_1$ and $c_3$, respectively—representing the right-most labels for those nodes.

Overall, Algorithm~\ref{algo_ffd} and Algorithm~\ref{algo_dfd} can be seen as the process of recovering a single tree path.
We provide the complete algorithms in \cref{appendix:algos} for systematically reconstructing all paths.

\subsection{Discussion and Potential Optimizations}

\subsubsection{Different Paths with Identical Labels}
\label{sec:discuss_same_label}
As claimed in Section~\ref{sec:model rep}, we adopt the same assumption as \cite{tramer2016stealing}, where each path (\ie leaf) is assigned a unique identifier, allowing the attacker to distinguish between different tree paths by feeding the \ac{DT} with customized inputs.
This assumption makes regression trees particularly well-suited for \sys, as each leaf node has a unique real-valued label. However, in classification trees, both \cite{tramer2016stealing} and \sys may fail to recover some paths if multiple paths share the same label. 
To address this limitation, we propose the following solutions for paths with:

\noindent\textbf{Same label but different node counts.} 
Under our threat model (\cf Section~\ref{sec:threat model}), \sys can distinguish such paths based on node counts, similar to \cite{tramer2016stealing}.

\noindent\textbf{Same label and same node counts.} 
If an attacker has the capability to measure fine-grained timing variations, it may be possible to differentiate paths based on total inference time. 
The left and right branches at each node may introduce slight timing variations due to jump instructions, and if two paths follow different branching patterns, their overall execution time will differ, allowing for differentiation.
However, if two paths not only share the same label and node count but also exhibit identical branching patterns (\ie the same number of left and right branches), this approach becomes ineffective.
A potential alternative is to leverage the confidence values associated with different leaves to distinguish them (\cf Section~\ref{sec:dt basics}). 
However, this method may not be effective in all cases and requires further investigation.

\subsubsection{Complexity Analysis}
\label{sec:complexity}
Since \cite{tramer2016stealing} is the most relevant work under similar assumptions, we compare \sys against it.  
Let $m$, $d$, $h$, and $\epsilon$ represent the number of leaves in the tree, the number of continuous features, the tree depth, and the granularity used in binary search, respectively, where $m = 2^{h-1}$. 
To analyze complexity, we adopt the same assumption that each leaf is assigned a unique identifier and that no continuous feature is split into intervals smaller than $\epsilon$. 
For continuous features within the range $[0, b]$, determining a single threshold requires at most $\log_2{\frac{b}{\epsilon}}$ queries. 
The complexity of \cite{tramer2016stealing} is $O(m^2 \cdot d \cdot \log_2{\frac{b}{\epsilon}})$.  
For \sys, in the worst case (without distinguishing between unique and duplicate features), the complexity is $O(m \cdot (d + 2 \cdot d \cdot h + d \cdot \log_2{\frac{b}{\epsilon}} + 2 \cdot d)) \approx O(m \cdot d \cdot (2h + \log_2{\frac{b}{\epsilon}}))$.
\sys achieves at least one order of magnitude improvement over \cite{tramer2016stealing}.  
Furthermore, \sys incurs different overheads when handling unique and duplicate features along each path (see Section~\ref{sec:exp_breakdown} for breakdown benchmarks). 
As a result, when applied to real-world \ac{DT} models, our performance is expected to be significantly better than the worst-case bound.

\subsubsection{Stride-based Linear Search}
As shown in Section~\ref{sec:complexity}, binary search constitutes a significant portion of the overhead.
A potential alternative is to employ a stride-based linear search instead of relying solely on binary search in Algorithms~\ref{algo_bs}, \ref{algo_ffd}, and \ref{algo_dfd}.
Specifically, in the left subtree, the search begins at the lower bound, incrementally increasing the value by a \textit{carefully chosen stride} to locate the threshold. 
In the right subtree, starting from the upper bound and decreasing by the same stride.
This method avoids unnecessary searches from significantly larger bounds in binary search, reducing the number of queries required. 
However, the attacker must have an approximate understanding of the threshold distribution within the feature value range.
A key challenge lies in selecting an appropriate stride size—if the stride is too large, it may skip over the correct threshold. 
One possible solution is to first perform a linear search with a moderate stride to identify a smaller range containing the threshold, followed by a binary search within this refined range for precise recovery.
This approach is particularly beneficial for features with large value ranges, where binary search alone may be inefficient. 
We leave this interesting extension for future work.

\section{Experiments}
We implement a proof-of-concept attack for \sys in C, with a codebase of approximately 500 LoC.
Using this implementation, we first compare \sys with prior work. 
Next, we provide a detailed performance breakdown of \sys across different metrics using customized flexible \acp{DT}. 
Finally, we describe our experimental setup for fault injection via voltage glitching and present the results of applying real glitch attack on \ac{DT}.

\subsection{Datasets and Models}
To evaluate the performance of \sys, we utilize five commonly used datasets from the UCI Machine Learning Repository \cite{asuncion2007uci}. Detailed information about these datasets is provided in \autoref{table:cmp_uci_models}.
We adopt tree depths commonly trained for each dataset based on in prior work \cite{tramer2016stealing}.
For inference, we deploy \ac{DT} models on embedded devices. 
These models are generated using \ac{ML} library, Emlearn~\cite{emlearn}, that supports microcontrollers and embedded systems. 
Emlearn allows models to be trained in Python and then converted into a C classifier, enabling efficient inference on any device equipped with a C99 compiler.

\subsection{Comparison of \sys and Prior Work}
\label{sec:cmp_prior}

\begin{table*}[htbp]
\centering
\caption{\textbf{Performance of \sys on public models}. For each model, we report the number of features used in this model, the number of classes ($\mathbb{R}$ represents this tree is the regression tree), the number of leaves (\ie paths), the maximum tree depth, and $\epsilon$ denotes the chosen granularity. For results of \cite{tramer2016stealing}, \textit{Normal} denotes the extraction under normal assumptions. \textit{Top-down} denotes a stronger setting that allows querying the model with partially specified inputs, a capability supported only by a few platforms such as BigML and not representative of typical deployments. \textit{Fault Runs} denotes the required fault runs (already included in total), representing the theoretical lower bound assuming a 100\% fault-success rate.  In practice, the actual number varies across different injection techniques.}
\label{table:cmp_uci_models}
\resizebox{!}{1.5cm}{
\begin{tabular}{llllllllll}
\hline
\multirow{3}{*}{\textbf{Model}} & \multirow{3}{*}{\textbf{\# Features}} & \multirow{3}{*}{\textbf{\# Classes}} & \multirow{3}{*}{\textbf{\# Leaves}} & \multirow{3}{*}{\textbf{Depth}} & \multirow{3}{*}{\textbf{$\epsilon$}} & \multicolumn{4}{l}{\textbf{\# Queries}} \\ \cline{7-10} 
 &  &  &  &  &  & \multicolumn{2}{l}{\cite{tramer2016stealing}} & \multicolumn{2}{l}{Ours} \\ \cline{7-10} 
 &  &  &  &  &  & Normal & Top-down & Total & Fault Runs \\ \hline
Iris & 3 & 3 & 9 & 5 & $10^{-3}$ & 243 & 169 & 164 & 45 \\
Diabetes Diagnosis & 5 & 2 & 11 & 5 & $10^{-3}$ & 662 & 256 & 248 & 29 \\ 
Medical Provider Charge & 11 & $\mathbb{R}$ & 50 & 10 & $10^{-3}$ & $10,092$ & $2,541$ & $1,718$ & 311 \\
Bitcoin Price & 6 & $\mathbb{R}$ & 147 & 11 & $10^{-4}$ & $23,315$ & $11,460$ & $4,092$ & 831 \\
Appliances Energy Prediction & 25 & $\mathbb{R}$ & 158 & 17 & $10^{-3}$ & $24,904$ & $10,342$ & $5,355$ & 329 \\ \hline
\end{tabular}
}
\end{table*}

As discussed in Section~\ref{sec:related_mea}, only a few studies~\cite{tramer2016stealing,chandrasekaran2020exploring,oksuz2024autolycus} have explored model extraction attacks on \ac{DT} models.
\cite{chandrasekaran2020exploring} formulates model extraction as query synthesis active learning. 
Their approach generally requires significantly more queries than \cite{tramer2016stealing}, but achieves higher accuracy.
Moreover, their results are currently not publicly available for reproduction.
\cite{oksuz2024autolycus} leverages \ac{XAI} technique for extracting \ac{DT} models. 
However, their method relies on rich auxiliary information, such as feature importance, which is also an attack goal in \sys. 
Among these, \cite{tramer2016stealing} is the most relevant to \sys, making it the primary baseline for our comparison. In the following, we present a detailed comparison between \sys and \cite{tramer2016stealing}.

\noindent\textbf{Attack evaluation metric.} 
To compare them, we primarily use the number of queries as the evaluation metric.
We do not include accuracy as a comparison metric because both \cite{tramer2016stealing} and \sys reconstruct models that achieve the same accuracy as the original model (\ie $1-R_{test}=100\%$ \footnote{This is the average error over the test set. A low test error indicates that the recovered \ac{DT} closely replicates the original model on inputs drawn from the same distribution as the training data.}). 
Specifically, we focus on models where each leaf (\ie path) has a unique identifier, making regression trees the most suitable choice.
In this scenario, \sys produces a model equivalent to the original as long as an appropriate $\epsilon$ is selected. 
$\epsilon$ was set experimentally based on feature ranges, ensuring accuracy matches normal inference.
For consistency, we use the same $\epsilon$ values as in \cite{tramer2016stealing} for different datasets in our experiments.

\noindent\textbf{Testbed.} 
Although we use the same UCI datasets as listed in \cite{tramer2016stealing}, we do not directly report their published results. 
This is because, even when selecting the same datasets and tree depths, the total number of nodes and leaves in a BigML-generated tree cannot be controlled. 
Variations in tree structure ultimately lead to differences in the number of queries required for extraction.
To ensure a fair comparison, we first generate \acp{DT} using the datasets listed in \autoref{table:cmp_uci_models} through BigML’s online dashboard. 
We then convert the BigML-style tree into the Emlearn-compatible format. 
This transformation guarantees that the evaluated trees have an identical structure and number of nodes across experiments.

Since the code of \cite{tramer2016stealing} is no longer maintained and some dependencies have become outdated, we reconstructed their attack code within a Docker environment using compatible dependencies\footnote{We will publicly release this Docker image to facilitate reproducibility.}.
For running \sys, we use Ubuntu 24.04 on a Lenovo ThinkStation P2 Tower, equipped with 32 GB RAM and an Intel Core i7-14700 CPU (up to 5.30 GHz). The Docker container running \cite{tramer2016stealing}’s code is also executed on this machine.

\noindent\textbf{Results.} 
In our proof-of-concept attack for \sys, we report the total number of queries and the required fault runs in \autoref{table:cmp_uci_models}.
For Iris and Diabetes Diagnosis, although these datasets correspond to classification trees with only two or three classes, the paths leading to the same class label contain different numbers of nodes, allowing the attacker to differentiate them.
Note that we do not include results using the incomplete query functionality offered by BigML, as this feature is not commonly supported by other \ac{ML} services. 
Moreover, this feature is overly powerful—reveals the node's feature by feeding an incomplete query with the missing feature, making extraction much easier and less representative of real-world threat scenarios.

As expected in the complexity analysis (\cf Section~\ref{sec:complexity}), for smaller \acp{DT} such as Iris and Diabetes Diagnosis, \sys achieves only a slight improvement over \cite{tramer2016stealing} (\eg 164 vs. 243 queries). 
However, for larger \acp{DT}, \sys requires significantly fewer queries (\eg $4,092$ vs. $23, 315$ queries).
This gain is primarily due to \sys’s fault injection strategy that injects faults at target nodes and employs a bottom-up approach to iteratively refine the search space, minimizing unnecessary queries.
Although deploying \sys requires additional attacker capabilities, such as fault-injection equipment and basic side-channel monitoring, this overhead is consistent with standard practice in implementation-level attacks. 
Moreover, these capabilities can be relaxed with alternative injection techniques (\eg software-based or microarchitectural approaches).
Overall, \sys not only reduces query complexity but also extracts more information, such as the tree structure and duplicate features within a path, which are not recovered by prior methods. 

\subsection{Evaluating \sys under Different Conditions}
\label{sec:exp_breakdown}
\begin{figure}[htbp]
  \centering
    \begin{subfigure}{0.49\linewidth}
      \centering   
      \includegraphics[height=3.7cm,width=\linewidth]{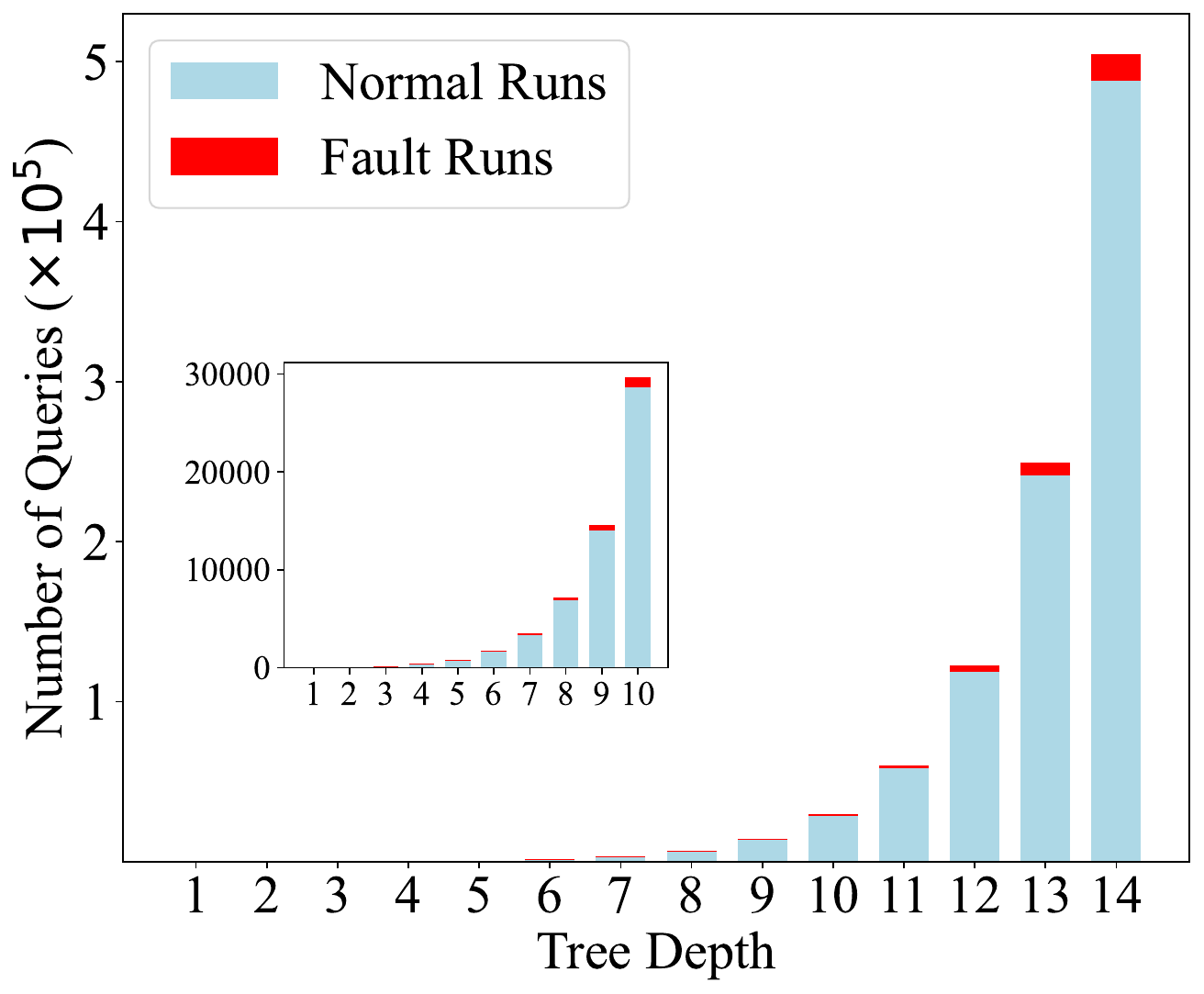}
        \caption{\scriptsize \#Features = 14 without duplicate features}
        \label{fig:breakdown:sub1}
    \end{subfigure}   
    \begin{subfigure}{0.49\linewidth}
      \centering   
      \includegraphics[height=3.7cm,width=\linewidth]{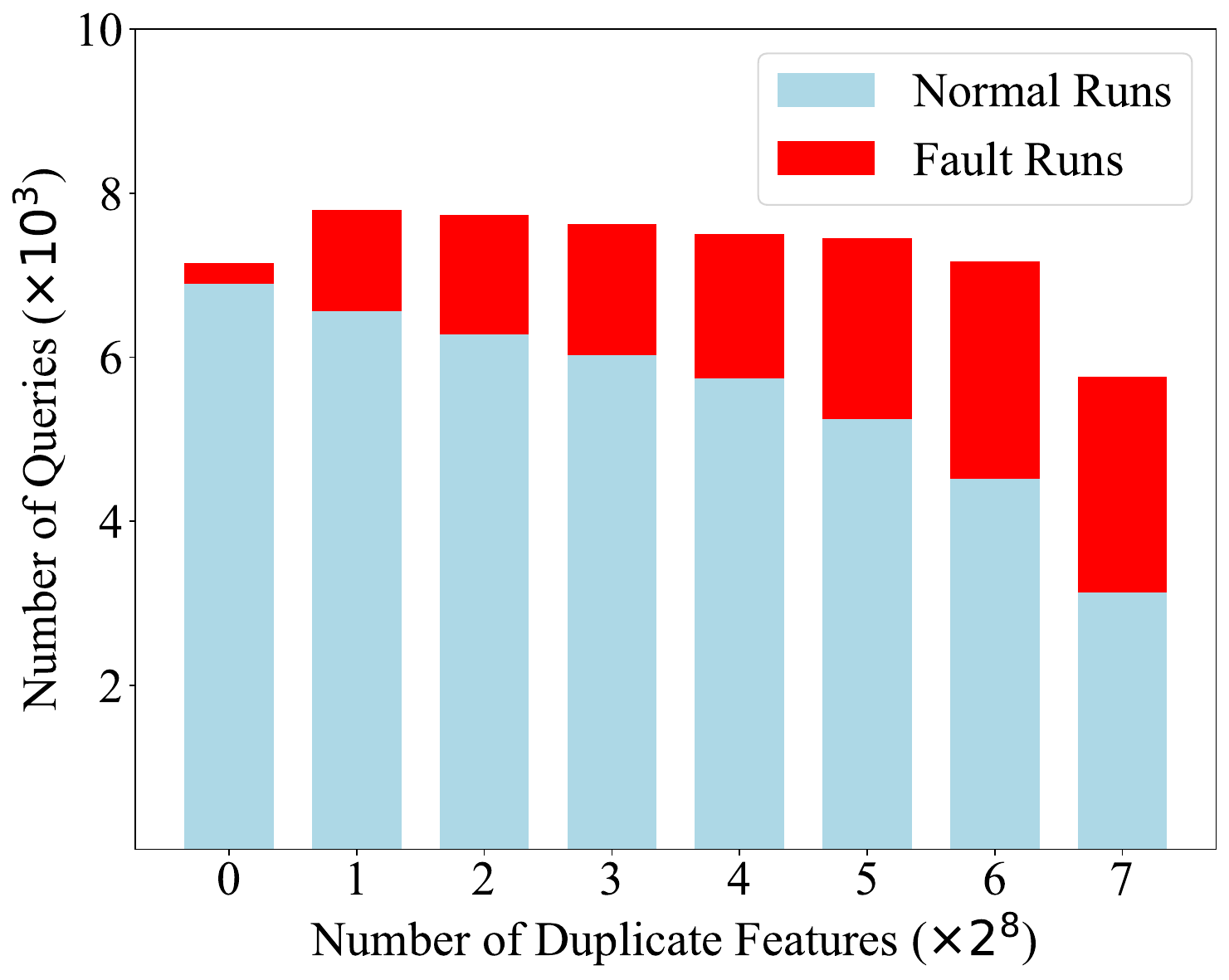}
        \caption{\scriptsize \#Depth = 8 and \#Features = 8}
        \label{fig:breakdown:sub2}
    \end{subfigure}
\caption{
\label{fig:breakdown}
Performance of \sys under varying tree depths (\autoref{fig:breakdown:sub1}, without duplicate features) and varying number of duplicate features (\autoref{fig:breakdown:sub2}, with depth fixed). $\epsilon=10^{-3}$.}
\end{figure}

Based on the comparison with prior work, we identify tree depth and the number of duplicate features as two key factors that influence the number of queries required for model extraction.
To systematically evaluate the impact of these factors, we assess the performance of \sys using synthetic \acp{DT}, which allow controlled variation in depth and feature duplication. 
These trees are generated using our custom Python script that produces complete binary trees with adjustable depth and numbers of duplicate features. 
The format of these trees is consistent with models generated by Emlearn.

In \autoref{fig:breakdown:sub1}, we fix the number of features to 14 and vary the tree depth from 1 to 14, ensuring no duplicate features on any path. 
The results show that as tree depth increases, both the total number of queries and fault runs grow exponentially. 
For example, the number of queries increases from $401$ at depth $4$ to $833$ at depth $5$. 
This trend is expected, as deeper trees contain exponentially more internal nodes and decision paths.

In \autoref{fig:breakdown:sub2}, we fix the tree depth to 8 and the number of features to 8, then increase the number of duplicate features from 0 to 7. 
We define duplicate features as follows: given depth $h$, a path with all $h$ nodes using the same feature results in $h - 1$ duplicate features; a path with distinct features has $0$ duplicates.
Interestingly, the trend in \autoref{fig:breakdown:sub2} does not follow a strictly increasing pattern. 
We observe the following:
\ding{182} As the number of duplicate features increases, the number of fault runs also rises, as expected, due to additional invocations of $\mathtt{DFD}()$. However, the number of queries increases sharply when moving from 0 to 1 duplicate feature and then grows only gradually thereafter. 
This is because, once \sys invokes $\mathtt{DFD}()$ for the first time, new paths are extended from the baseline path. 
This allows some previously recovered nodes to be reused, avoiding redundant recovery efforts. 
Furthermore, as duplicate features increase, the search space for subsequent thresholds becomes narrower, contributing to reduced query overhead.
\ding{183} When each path contains the maximum number of duplicate features (\ie 7), the total number of queries is moderately lower than other cases. 
This further supports the above observation.
In summary, these results highlight two key insights:
\circled{1} In \sys, an increase in duplicate features leads to more fault runs, which may impact overall efficiency—particularly in real-world scenarios where each fault injection may require multiple attempts (\cf Section~\ref{sec:glitch_exp} for practical results).
\circled{2} Despite the additional fault runs, the number of queries does not increase substantially with more duplicate features and, under certain conditions, may even decrease.

\subsection{Results of Voltage Glitch-induced Faults in Inference}
\label{sec:glitch_exp}
In this section, we demonstrate how practical faults can be induced in our proof-of-concept attack using voltage glitch, one of the most effective fault injection techniques.

\subsubsection{Hardware Setup}
\label{sec:hw_setup}

\noindent\textbf{Crowbar glitch mechanism.}
The glitching mechanism utilized in our setup is based on a crowbar circuit~\cite{van2021hardware}, which momentarily short-circuits the power rails of the target device. The resulting glitch waveform depends on the specific characteristics of the target’s power supply configuration.

\noindent\textbf{Glitch tool.}
For the voltage glitch attacks, we employ Faultier \cite{faultier} to generate the required glitch waveforms. 
Faultier integrates an N-Channel MOSFET (DMG2302UK) for crowbar glitching and includes a dual-channel analog switch for performing voltage dips or re-powering the target device. 
In addition, Faultier provides a basic \ac{SWD} probe, allowing the extraction of memory contents from glitched chips and verifying the success of glitching by re-enabling debugging features. 
The tool supports glitch resolution up to 200 MHz (5 ns), offering fine-grained control over glitch timing \footnote{\url{https://app.hextree.io/}}.

\begin{figure}[htbp]
\centering
\includegraphics[width=0.65\linewidth]{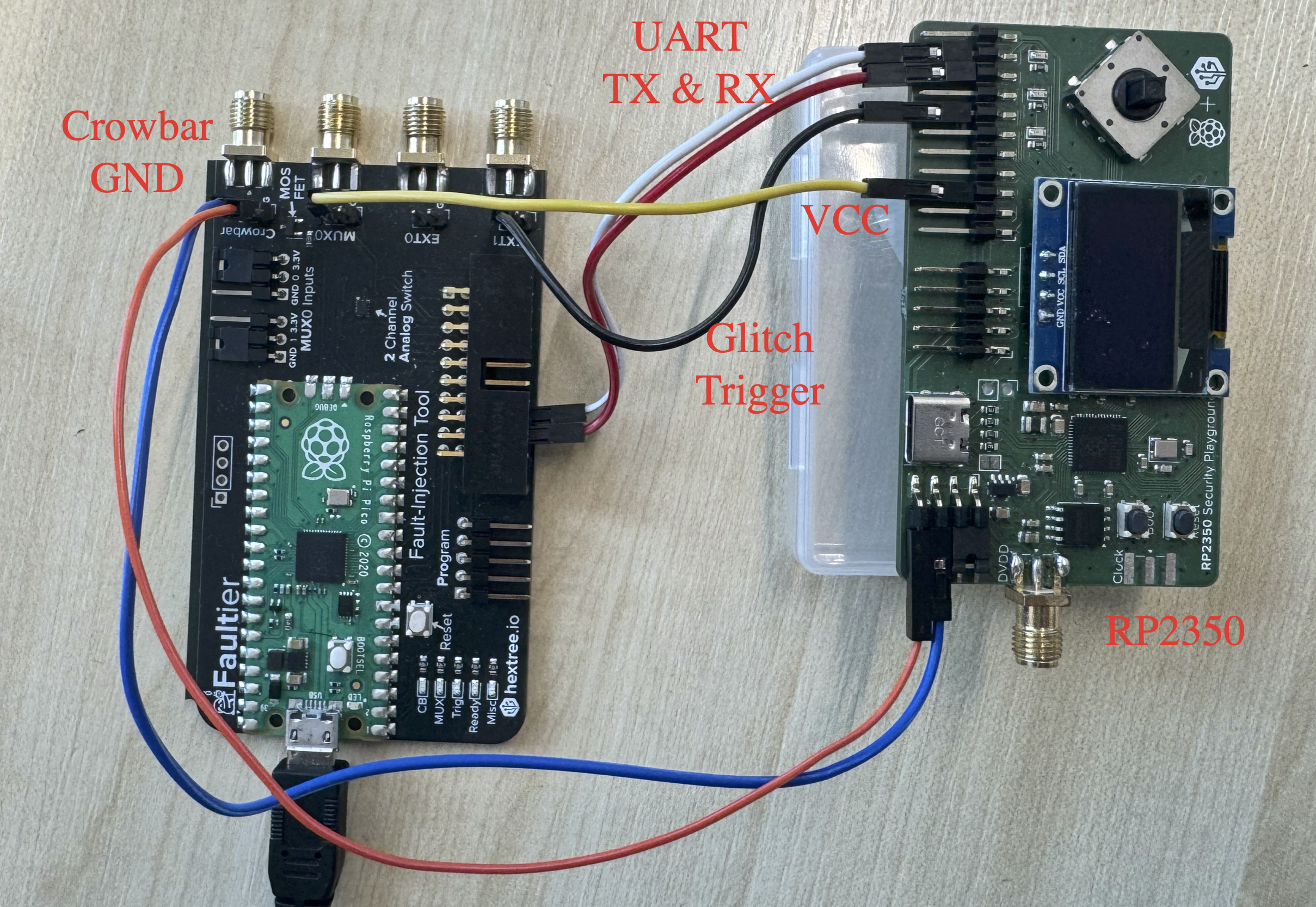}
\caption{Hardware setup using Faultier.} 
\label{fig:faultier}
\end{figure}

\begin{figure}[!t]
  \centering
    \begin{subfigure}{0.42\linewidth}
      \centering   
      \includegraphics[height=2.7cm, keepaspectratio]{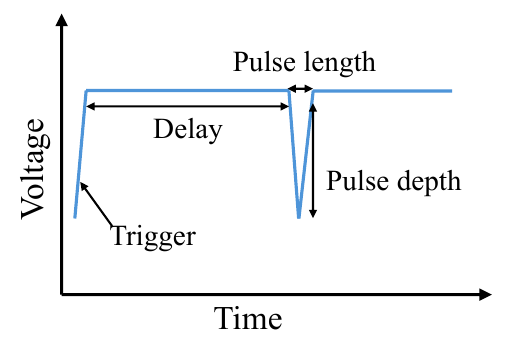}
        \caption{}
        \label{fig:glitch_params:sub1}
    \end{subfigure}   
    \begin{subfigure}{0.49\linewidth}
      \centering   
      \includegraphics[height=2.7cm, keepaspectratio]{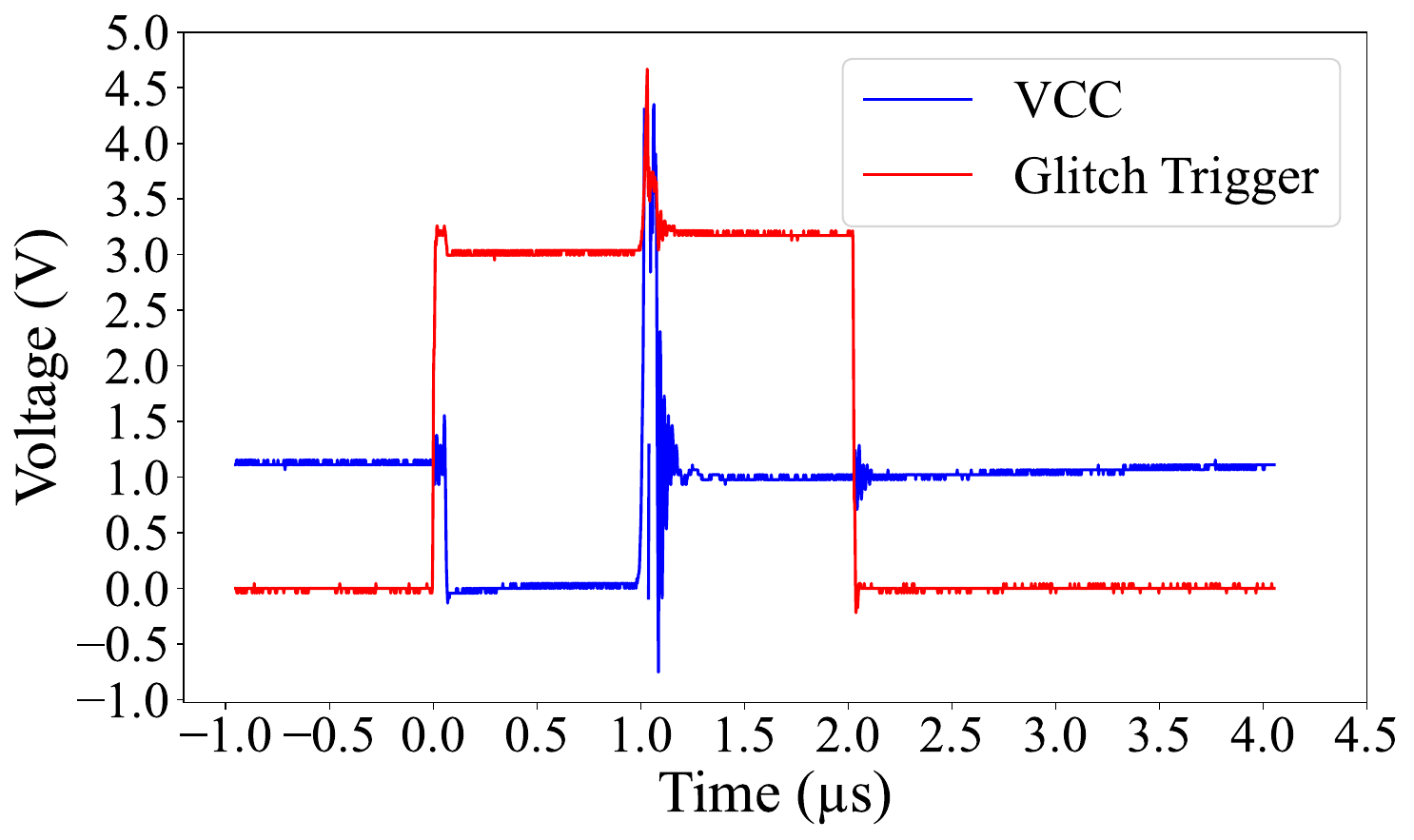}
        \caption{}
        \label{fig:glitch_params:sub2}
    \end{subfigure}
\caption{
\label{fig:glitch_params}
Glitch parameter conventions used in this experiment (\autoref{fig:glitch_params:sub1}) and waveforms for a successful glitch, including VCC and glitch trigger signals (\autoref{fig:glitch_params:sub2}).}
\end{figure}

\noindent\textbf{Target device.} 
We use the RP2350 Security Playground board, developed by the Faultier team, as shown in \autoref{fig:faultier}. 
This platform is specifically designed to evaluate fault injection resilience on RP2350-based systems. 
It includes built-in support for bypassing glitch detection and performing OTP fault testing, making it well-suited for practical security experiments. 
In our setup, we run \ac{DT} inference on the board with security features like glitch detection deliberately turned off.

\subsubsection{Signals during Fault Injections}
As shown in \autoref{fig:glitch_params:sub1}, two key parameters define the glitch configuration: the glitch delay (\texttt{delay}), representing the time between the trigger signal and the glitch; and the glitch pulse length (\texttt{pulse}), indicating the duration of the glitch. 
During glitching attempts, we primarily adjust the (\texttt{delay}, \texttt{pulse}) pair (in cycle length) in Faultier to achieve successful fault injection on the target.
\autoref{fig:glitch_params:sub2} shows the waveforms of a successful glitch at a target node, where the injected fault forces the branch to take the left path.

\subsubsection{Glitch Results for Inference}

\begin{figure}[htbp]
\centering
\includegraphics[width=0.45\linewidth]{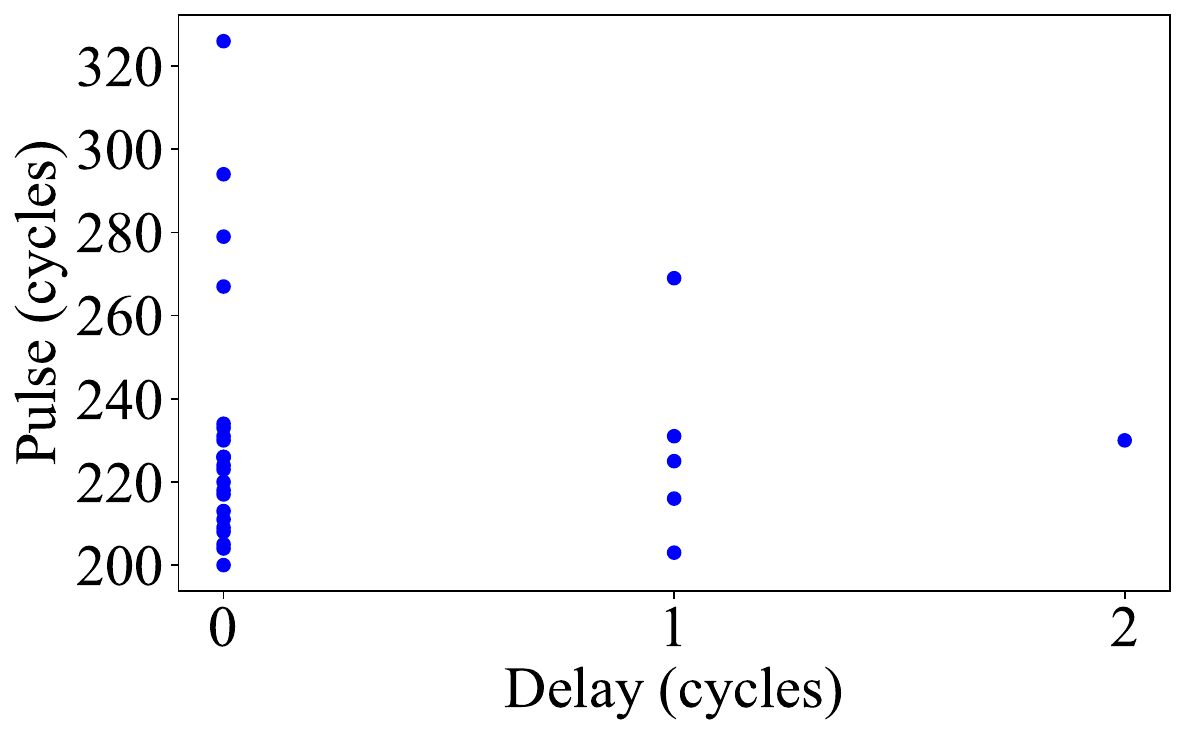}
\caption{Successful glitch parameter pairs, where each blue dot corresponds to a single (delay, pulse) pair that induced a valid fault.} 
\label{fig:real_glitch}
\end{figure}

To evaluate our attack with the voltage glitch, we fine-tune the glitch parameter pair to induce the desired fault type (\eg forcing the branch to go left or right). 
Accordingly, we focus on the fault runs reported in \autoref{table:cmp_uci_models}, using the tree trained on the Diabetes Diagnosis dataset as a representative example.
For each of the 29 fault runs, we record the first successful (\texttt{delay}, \texttt{pulse}) pair that induces a valid fault, with results presented in ~\autoref{fig:real_glitch}. 
The glitching process involves gradually increasing both delay and pulse values from zero while monitoring the system response to detect a successful injection.
In principle, we can define large search ranges—\eg $delay \in [0, 1000]$ and $pulse \in [0, 500]$ (as done in Faultier's tutorials \cite{faultier})—to ensure the space includes at least one successful pair. However, a more efficient strategy is to first perform offline glitching on a structurally similar tree to identify a viable glitch pair for each node. 
Based on this empirical knowledge, we can significantly narrow the search range when faulting one node. For example, given a successful glitch pair $(1, 224)$, we can restrict the search to delay $delay \in [1, 10]$ and $pulse \in [200, 250]$, thereby avoiding unnecessary attempts.
In total, 703 extra runs were required—higher than in our idealized proof-of-concept—resulting in an overall execution time of approximately $70,994$ cycles, say $0.53$ ms at a 133 MHz CPU frequency. 
This overhead is expected due to the inherently probabilistic nature of voltage glitching and its limited precision.
Nevertheless, in this evaluation, voltage glitching primarily serves as a practical demonstration of our attack methodology. 
We believe that employing more precise and controllable fault injection techniques—such as EM or laser-based methods—could significantly reduce the gap between the real-world performance of \sys and its ideal implementation.

\section{Limitations and Discussions on More Use Cases}
\label{sec:limitation_discussion}
\subsection{Limitations in Fault Injection Efficiency}
As shown in Section~\ref{sec:glitch_exp}, injecting faults at specific nodes via voltage glitching often requires multiple attempts for a single successful execution. In cases of improper setup or parameter tuning, the number of queries needed to recover the \ac{DT} can increase significantly.
To alleviate this, exploring alternative fault injection techniques for attacking \ac{DT} models could be valuable, especially in cloud scenarios instead of embedded devices. 
One potential approach is incorporating Rowhammer attacks into \sys, leveraging a hybrid offline-online workflow.
For instance, when testing on a system equipped with a 4GB DDR3 DIMM memory subsystem in either single- or dual-channel configurations, \cite{rakin2022deepsteal} indicate that Rowhammer attacks targeting pages with at least one MSB bit offset require hammering approximately $3,000$ rows per round, with an average execution time of 200 seconds.
Approximately, considering the timing constraints of Rowhammer and \sys, a possible strategy is to pre-hammer target memory rows offline while ensuring the tree is loaded into the vulnerable memory area before the estimated success window. 
However, for newer DRAM architectures such as DDR5, achieving a successful fault injection may take considerably longer, introducing further challenges.
Overall, investigating the feasibility and efficiency of alternative fault injection techniques within \sys remains an interesting direction for future research.

\subsection{Countermeasures}
As \sys integrates both model extraction and \acp{FIA}, countermeasures should be considered from each perspective.
From the perspective of model extraction, differential privacy has been suggested as a defense mechanism in prior studies~\cite{tramer2016stealing,oksuz2024autolycus}. 
However, typical strategies—such as adding noise to training data, employing static or dynamic output distortion, or perturbing feature importance scores—have demonstrated limited effectiveness against more sophisticated extraction methods. 
This underscores the necessity of investigating advanced, adaptive differential privacy solutions.
From the perspective of fault injection, hardware-based protections~\cite{barenghi2012fault}, such as glitch detectors, voltage and clock monitors, and fast brown-out protection, can detect abnormal operating conditions and abort execution, thereby reducing the reliability of \sys. However, these defenses are not universally deployed on lightweight IoT and embedded platforms, and prior work shows that well-timed or low-amplitude glitches can sometimes bypass such monitors, meaning \sys may still be feasible in practice depending on the device’s hardware hardening.
Additionally, redundancy-based methods, such as duplicating critical computations combined with majority voting or result comparison, can detect discrepancies caused by faults, thereby reducing their impact.

\subsection{Extension to Other Tree Variants}

\noindent\textbf{GBDT/XGBoost.} 
\ac{GBDT} \cite{friedman2001greedy} and its optimized variant, XGBoost \cite{chen2016xgboost}, are ensemble learning methods that iteratively train shallow \acp{DT}, where each tree corrects the residual errors of the previous ones. Unlike traditional \acp{DT}, GBDT and XGBoost aggregate predictions from multiple trees using gradient-based optimization and regularization to improve accuracy and prevent overfitting.
Due to this aggregation mechanism, directly applying \sys to recover individual nodes in GBDT/XGBoost is impractical, as only the final prediction is exposed.
However, inspired by previous fault injection attacks—such as those on AES to infer the last-round key~\cite{tunstall2011differential} and on \acp{NN} to extract weights and biases~\cite{breier2021sniff}—a similar approach can be explored. 
By strategically injecting faults at specific nodes and analyzing the resulting variations in the final output, it may be possible to reconstruct node parameters by leveraging the known output formula of GBDT/XGBoost.

\noindent\textbf{Hoeffding tree in data stream classification.} 
The Hoeffding Tree (HT), a variant of the \ac{DT}, is widely used for data stream classification due to its ability to continuously and efficiently learn from incoming data \cite{domingos2000mining}. 
Unlike traditional \acp{DT}, which are trained on static datasets, HTs are incrementally updated on-the-fly as new data arrives.
A key characteristic of HT is that it can be queried at any time to provide a prediction. 
However, given the same input, the returned label may vary because the model dynamically evolves, with nodes being updated or even pruned based on a concept drift detector that identifies significant accuracy degradation.
Applying \sys to HTs enables the recovery of modified nodes or subtrees. 
Since changes in the HT structure primarily reflect shifts in the data stream, a potential research direction is to explore whether these recovered components can help an attacker infer the underlying data distribution of the stream. 

\subsection{Tree-based Cryptographic Schemes}
\label{sec:discuss_tree_crypto}
\acp{DT} have also found widespread use in cryptographic applications. For example, ZKFault~\cite{mondal2024zkfault} recently demonstrated that fault attacks on zero-knowledge-based post-quantum digital signature schemes can be devastating—allowing full secret key recovery in LESS and CROSS from a single fault injected into the seed tree.
\sys provides a complementary approach by physically injecting faults at specific nodes via voltage glitching. A promising future direction is to explore whether this targeted fault methodology can also compromise seed trees or other tree-based structures in cryptographic protocols, potentially revealing secret values.

For instance, function secret sharing (FSS) protocols such as distributed point functions (DPFs)~\cite{boyle2016function} encode functions as binary trees split between multiple parties. If a fault is injected into one party’s decision path—\eg flipping a comparison result—it may lead to selecting an incorrect leaf value (commonly used in \textit{oblivious array access} \cite{doerner2017scaling}).
Originally, a 2PC DPF protocol would return $a_i$ (the i-th element of array $a$); after a fault, the result might become $a_j$.
When this is explored in \ac{ML} domains, the incorrect selection of $a_j$ will lead to different labels (or say logits in \acp{NN}).
It is interesting to investigate if we can leverage this indirect leakage to infer internal parameters of \ac{ML} models.

Moreover, even privacy-preserving \ac{DT} frameworks, such as those surveyed in~\cite{kiss2019sok}, which use homomorphic encryption, secret sharing, or \acp{TEE}, could be vulnerable to fault injection. For example, if a malicious client in such a system can inject transient faults into the model-serving enclave or MPC server and observe changes in the prediction results (as done in \sys), they may be able to infer the underlying tree structure or internal thresholds. This creates a new dimension of side-channel leakage that bypasses cryptographic protections and underscores the need for robust hardware-level fault countermeasures.

\section{Conclusion}
In this work, we introduce \sys, a novel model stealing attack that leverages fault injection to recover tree models. 
We design a bottom-up recovery algorithm combined with targeted fault injection at specific nodes, which significantly improves the efficiency of the search process for feature thresholds, thereby reducing the number of required queries.
Our approach is capable of recovering all internal nodes and their corresponding threshold values. 
With a properly chosen search granularity $\epsilon$, \sys can reconstruct a model that closely matches the original one.
Through our proof-of-concept implementation, we demonstrate that \sys not only reduces query complexity compared to prior work but also recovers richer structural information. 
Additionally, we validate the practicality of our approach by deploying \sys on a microcontroller using the voltage glitch tool Faultier, successfully performing real-world fault injections to extract model parameters.
Since \sys targets general \ac{DT} models, it holds strong potential for broader applications. We further discuss its applicability to other domains, including popular tree-based models such as GBDT, XGBoost, and Hoeffding Trees, as well as tree-based cryptographic schemes.

\begin{acks}
We would like to thank Thomas ``stacksmashing'' Roth of hextree.io for providing us with early access to the Faultier hardware platform. This work has been supported by the BMFTR through the projects AnoMed and SILGENTAS.
This research is partially funded by the Engineering and Physical Sciences Research Council (EPSRC) under grant EP/X03738X/1.
\end{acks}

\bibliographystyle{ACM-Reference-Format}
\bibliography{ref}

\clearpage
\appendix

\section{Complete Procedure of \sys}
\label{appendix:algos}

Algorithm~\ref{algo_ffd} and Algorithm~\ref{algo_dfd} provide the details of the procedure of discovering first occurrence of features (Section~\ref{sec:ffd}) and discovering duplicate features (Section~\ref{sec:dfd}), respectively. 

Algorithm~\ref{algo_rti} and Algorithm~\ref{algo_general} serve as iterators to systematically reconstruct all paths in a \ac{DT}.
To iteratively recover paths based on previously recovered ones, a crucial step is generating new input arrays that will traverse along unexplored paths. 
Given $j$-th node (\ie $x_i<t$), we modify $\bm{X}[i]$ to $t \pm \epsilon$, ensuring that the modified input leads to a new label where branching occurs at the $j$-th node.(Lines~\ref{algo_rti:13}-\ref{algo_rti:19} in Algorithm~\ref{algo_rti}).
As outlined in Algorithm~\ref{algo_general}, the recovery process begins by selecting the leftmost and rightmost paths as the \textit{baseline paths}, which are then used to systematically reconstruct all other \textit{extended} paths.

\begin{algorithm}[]
\renewcommand{\algorithmicrequire}{\textbf{Input:}}
\renewcommand{\algorithmicensure}{\textbf{Output:}}
\footnotesize
\caption{First Feature Discovery: $\mathtt{FFD}()$}
\label{algo_ffd}
\begin{algorithmic}[1]
\Require Path $\bm{P}$, input $\bm{X}$, timing array $\mathcal{T}$ of $\bm{P}$, left/right subtree flag $flag$, duplicate feature candidate set $S_{DF}$, granularity $\epsilon$
\Ensure Updated $\bm{P}$ and $S_{DF}$

\State Initialize array $S_{DF}$ which stores the duplicate features on path $\bm{P}$

\For{$i = 0$ to $d-1$}
    \State $\bm{X'} \gets \bm{X}$
    \State $\bm{X'}[i] \gets (flag = 0)\ ?\ (s_i.b+1)\ :\ (s_i.a-1)$ \Comment{Change a feature value to \textit{Min/Max}}
    \State $\bm{P'}.c \gets \mathtt{Inf}(\bm{X'})$ \Comment{$i$-th feature is on path $\bm{P}$ if $\bm{P'}.c \ne \bm{P}.c$}
    \State $\bm{P'}, \mathcal{T'} \gets \mathtt{CTN}(\zeta_{sc}, \bm{P'})$

    \If{$\bm{P'}.c \ne \bm{P}.c$}
        \For{$j = 0$ to $\bm{P'}.\beta - 1$}
            \State $c_f \gets \mathtt{F\_Inf}(\bm{X'}, \mathcal{T'}, j, flag)$
            
            \If{$c_f \ne \bm{P'}.c$} \Comment{Indicate $\mathtt{F\_Inf}()$ alters the result of the $i$-th feature at $j$-th node}
                \State $\bm{P}.\bm{V_j}.s \gets s_{i}$
                
                \If{$c_f \ne \bm{P}.c$} \Comment{Indicate it traversed a new branch, thus $s_i$ is duplicated}
                    \State $S_{DF}.append(s_{i})$
                \Else
                    \State $\bm{P}.\bm{V_j}.t \gets \mathtt{FaBS}(\bm{X'}, \bm{P}.c, \mathcal{T}, i, \textbf{None}, s_i.a, s_i.b, \textbf{False}, flag, \epsilon)$ \Comment{Unique feature $s_i$}
                \EndIf
                \State \textbf{break}
            \EndIf
        \EndFor
    \EndIf
\EndFor
\end{algorithmic}
\end{algorithm}

\begin{algorithm}[]
\renewcommand{\algorithmicrequire}{\textbf{Input:}}
\renewcommand{\algorithmicensure}{\textbf{Output:}}
\footnotesize
\caption{Duplicate Features Discovery: $\mathtt{DFD}()$}
\label{algo_dfd}
\begin{algorithmic}[1]
\Require Path $\bm{P}$, input $\bm{X}$, timing array $\mathcal{T}$ of $\bm{P}$, array $fea\_range$ records max/min value for each feature on this path, left/right subtree flag $flag$, duplicate feature array $S_{DF}$, granularity $\epsilon$
\Ensure Updated $\bm{P}$

\State $tidx \gets 0$ \Comment{Indicate the current threshold's index}
\State Initialize arrays $c_b$ and $loc$ of size $d$ which store the baseline label and node index for each feature, respectively
\State Initialize array $\bm{t}$ of size $d \times h$ (depth $h$), which stores each feature's previously confirmed threshold

\For{$s_i \in S_{DF}$} \Comment{Search each duplicate feature's first threshold value (the one closest to the leaf)}
    \State $c_b[i] \gets \bm{P}.c$
    \State $\bm{t}[i][tidx] \gets \mathtt{FaBS}(\bm{X}, c_b[i], \mathcal{T}, i, None, fea\_range[i][0], \allowbreak fea\_range[i][1], False, flag, \epsilon)$
\EndFor

\While{not all $\bm{P}.\bm{V_j}.t$ are recovered}
    \For{$s_i \in S_{DF}$} \Comment{\ding{182} Identify the node containing the discovered threshold}
        \State $\bm{X'} \gets \bm{X}$
        \State $\bm{X'}[i] \gets (flag = 0)\ ?\ \bm{t}[i][tidx] + \epsilon : \bm{t}[i][tidx] - \epsilon$
        \For{$j = \bm{P}.\beta-1$ to $0$}
            \If{$\bm{P}.\bm{V_j}.t \ne null$}
                \State \textbf{continue} \Comment{Node is already recovered}
            \EndIf
            \If{$\bm{P}.\bm{V_j}.s = s_i$}
                \State $\bm{P}.\bm{V_j}.t \gets tres[i][tidx]$
                \State $S_{DF}.remove(s_i)$ \Comment{Last node containing $s_i$ has been confirmed}
                \State \textbf{break}
            \EndIf
            \State $c_f \gets \mathtt{F\_Inf}(\bm{X'}, \mathcal{T}, j, flag)$
            \If{$c_f = c_b$}
                \State $\bm{P}.\bm{V_j}.s \gets s_i$
                \State $\bm{P}.\bm{V_j}.t \gets t_i^{tidx}$
                \State $loc[i] \gets j$ \Comment{Record the index of the previous node including $s_i$}
                \State \textbf{break}
            \EndIf
        \EndFor
    \EndFor

    \For{$s_i \in S_{DF}$} \Comment{\ding{183} Update feature value range and check if $s_i$ appears in the remaining nodes}
        \State $fea\_range[i][flag] \gets \bm{t}[i][tidx]$
        \State $\bm{X'}[i] \gets (flag = 0)\ ?\ fea\_range[i][1] : fea\_range[i][0]$
        \State $c_{ifdup} \gets \mathtt{F\_Inf}(\bm{X'}, \mathcal{T}, loc[i], flag)$
        \If{$c_{ifdup} = c_b[i]$}
            \State $S_{DF}.remove(s_i)$ \Comment{$s_i$ will not appear in the following nodes}
        \EndIf
    \EndFor

    \For{$s_i \in S_{DF}$} \Comment{\ding{184} Search for each duplicate feature's next threshold}
        \State $\bm{t}[i][tidx+1] \gets \mathtt{FaBS}(\bm{X}, c_b[i], \mathcal{T}, i, loc[i], fea\_range[i][0], \allowbreak fea\_range[i][1], True, flag, \epsilon)$
        \State $\bm{X'}[i] \gets (flag = 0)\ ?\ \bm{t}[i][tidx+1] - \epsilon : \bm{t}[i][tidx+1] + \epsilon$
        \State $c_b[i] \gets \mathtt{Inf}(\bm{X'})$
    \EndFor

    \State $tidx++$
\EndWhile
\end{algorithmic}
\end{algorithm}

\clearpage

\begin{algorithm}[]
\renewcommand{\algorithmicrequire}{\textbf{Input:}}
\renewcommand{\algorithmicensure}{\textbf{Output:}}
\scriptsize
\caption{Recover Tree Iterator: $\mathtt{RTI}()$}
\label{algo_rti}
\begin{algorithmic}[1]
\Require All paths $(\bm{P_0},\bm{P_1},{\cdots},\bm{P_{\alpha-1}})$ and their inputs $\bm{\overline{X}}$ of size ${\alpha}{\times}d$, left/right subtree flag $LR\_path$ for each $\bm{P}$, array $paths\_status$ tracking path discovery completion, array $start\_node$ specifying recovery starting nodes, array $candidates$ selecting baseline paths for extending to news, granularity $\epsilon$
\Ensure Updated $(\bm{P_0},\bm{P_1},{\cdots},\bm{P_{\alpha-1}})$

\For{$k = 0$ to $\alpha - 1$}
    \If{$candidates[k] \ne 1$}
        \State \textbf{continue}
    \EndIf

    \State Initialize arrays $\bm{t_{cur}}$, $\bm{br_{cur}}$ of size $d$ \Comment{Record identified threshold and branch direction in $\bm{P_k}$}

    \For{$i \gets start\_node[k]$ to $\bm{P_k}.\beta$} 
        \State $cur\_fea \gets \bm{P_k}.\bm{V_i}.s$; $cur\_br \gets \bm{P_k}.\bm{V_i}.br$; $cur\_t \gets \bm{P_k}.\bm{V_i}.t$

        \For{$j \gets 0$ to $i-1$}
            \If{$\bm{P_k}.\bm{V_j}.s = cur\_fea$} \Comment{Record the nearest node containing feature $cur\_fea$}
                \State $\bm{t_{cur}}[cur\_fea] \gets \bm{P_k}.\bm{V_j}.t$; $\bm{br_{cur}}[cur\_fea] \gets \bm{P_k}.\bm{V_j}.br$ 
            \EndIf
        \EndFor
        \Statex \Comment{Modify $cur\_fea$-th feature value in $\bm{\overline{X}}[k]$ to traverse a new path branching at $i$-th node in $\bm{P_k}$ $\quad$}
        \If{$\bm{br_{cur}}[cur\_fea] = null$} \Comment{Case: empty node} \label{algo_rti:13}
            \State $\bm{\overline{X}}[k][cur\_fea] \gets (cur\_br = 0)\ ?\ (cur\_t + \epsilon)\ :\ (cur\_t - \epsilon)$
        \ElsIf{$cur\_br = 0$} \Comment{Case: prev. branch is left/right and cur. is left}
            \State $\bm{\overline{X}}[k][cur\_fea] \gets cur\_t + \epsilon$
        \ElsIf{$cur\_br = 1$} \Comment{Case: prev. branch is left/right and cur. is right}
            \State $\bm{\overline{X}}[k][cur\_fea] \gets cur\_t - \epsilon$ 
        \EndIf \label{algo_rti:19}

        \State $c_m \gets \mathtt{Inf}(\bm{\overline{X}}[k])$ where $m \in (0, \alpha-1)$; $\bm{P_m}, \mathcal{T}_m \gets \mathtt{CTN}(\zeta_{sc}, \bm{P_m})$
        \State Initialize $fea\_range$ and $S_{DF}$

        \For{$j \gets 0$ to $\bm{P_m}.\beta - 1$} \Comment{Copy the information of the same nodes into the new $m$-th path}
            \If{$j < i + 1$}
                \State $\bm{P_m}.\bm{V_j}.s \gets \bm{P_k}.\bm{V_j}.s$
                \State $\bm{P_m}.\bm{V_j}.t \gets \bm{P_k}.\bm{V_j}.t$
                \State $\bm{P_m}.\bm{V_j}.br \gets (j \ne i)\ ?\ \bm{P_k}.\bm{V_j}.br\ :\ (1 - \bm{P_k}.\bm{V_j}.br)$
                \State $fea\_range[\bm{P_m}.\bm{V_j}.s][1 - \bm{P_m}.\bm{V_j}.br] \gets \bm{P_m}.\bm{V_j}.t$
                \State $S_{DF}.append(\bm{P_m}.\bm{V_j}.s)$
            \Else
                \State $\bm{P_m}.\bm{V_j}.br \gets LR\_path[k]$
            \EndIf
        \EndFor

        \For{$s_j \in S_{DF}$}  \Comment{Check if the dup. feature will appear in the remaining nodes on this path}
            \State $\bm{\overline{X}}[k][j] \gets (LR\_path[k]=0)\ ?\ (fea\_range[j][1] - \epsilon)\ :\ (fea\_range[j][0] + \epsilon)$
            \State $c \gets \mathtt{Inf}(\bm{\overline{X}}[k])$
            \If{$c = c_m$}
                \State $S_{DF}.remove(s_j)$
            \EndIf
        \EndFor

        \State $paths\_status[k] \gets 0$; $LR\_path[m] \gets LR\_path[k]$  \Comment{Update auxiliary information for next loop}

        \If{$\bm{P_m}.\beta = i + 1$}
            \State $candidates[k] \gets 0$
            \State \textbf{continue}
        \EndIf

        \State $\mathtt{FFD}(\bm{P_m}, \bm{\overline{X}}[k], \mathcal{T}_m, LR\_path[k], S_{DF}, \epsilon)$
        \State $\mathtt{DFD}(\bm{P_m}, \bm{\overline{X}}[k], \mathcal{T}_m, fea\_range, LR\_path[k], S_{DF}, \epsilon)$

        \State $candidates[k] \gets 0$; $candidates[m] \gets 1$; $\bm{\overline{X}}[m] \gets \bm{\overline{X}}[k]$; $start\_node[m] \gets i + 1$
    \EndFor
\EndFor
\end{algorithmic}
\end{algorithm}

\begin{algorithm}[]
\renewcommand{\algorithmicrequire}{\textbf{Input:}}
\renewcommand{\algorithmicensure}{\textbf{Output:}}
\scriptsize
\caption{\ac{DT} Extraction: $\mathtt{TreeExt}()$}
\label{algo_general}
\begin{algorithmic}[1]
\Require Features $\bm{S} \gets (s_0,s_1,{\cdots},s_{d-1})$
\Ensure Recovered tree $\Upsilon$

\State Generate $\bm{X_0}$ and $\bm{X_{\alpha - 1}}$, set $x_i \gets s_i.a - 1$ in $\bm{X_0}$ and $x_i \gets s_i.b + 1$ in $\bm{X_{\alpha - 1}}$ for $i \in [0, d-1]$

\State Initialize baseline input array $\bm{\overline{X}}$ of size ${\alpha}{\times}d$, where each $\bm{\overline{X}}[]$ represents an input designed to traverse a path
\State Initialize $LR\_path$ indicating whether each $\bm{P}$ is in the left or right subtree
\State Initialize $paths\_status$, tracking completion of each path
\State Initialize $start\_node$, the starting node index for recovery on each path
\State Initialize $candidates$, indicating which paths serve as baseline paths for extension in the loop
\State Initialize $S_{DF}$ and $fea\_range$ where $fea\_range[i][0]=s_i.a$ and $fea\_range[i][1]=s_i.b$
\State Define granularity $\epsilon$ used in $\mathtt{FaBS}()$, $\mathtt{FFD}()$, and $\mathtt{DFD}()$

\For{$i \in \{0, \alpha - 1\}$} \Comment{Identify the leftmost/rightmost paths as baselines for extending new paths}
    \State $c_i \gets \mathtt{Inf}(\bm{X_i})$
    \State $\bm{P_i}, \mathcal{T}_i \gets \mathtt{CTN}(\zeta_{sc}, \bm{P_i})$
    \State $\mathtt{FFD}(\bm{P_i}, \bm{X_i}, \mathcal{T}_i, i, S_{DF}, \epsilon)$
    \State $\mathtt{DFD}(\bm{P_i}, \bm{X_i}, \mathcal{T}_i, fea\_range, i, S_{DF}, \epsilon)$
    \State $paths\_status[i] \gets 0$; $start\_node[i] \gets 1$; $candidates[i] \gets 1$; $\bm{\overline{X}}[i] \gets \bm{X_i}$; $LR\_path[i] \gets i$ 
\EndFor

\While{$\mathtt{Sum}(paths\_status) \ne 0$} 
    \State $\mathtt{RTI}((\bm{P_0},\bm{P_1},{\cdots},\bm{P_{\alpha-1}}), \bm{\overline{X}}, LR\_path, paths\_status, start\_node, candidates, \epsilon)$
\EndWhile

\State \Return $\Upsilon \gets (\bm{P_0},\bm{P_1},{\cdots},\bm{P_{\alpha-1}})$
\end{algorithmic}
\end{algorithm}

\end{document}